\documentclass[aps,prd,preprint,groupedaddress,preprintnumbers,longbibliography,nofootinbib]{revtex4-1}

\usepackage{amsmath}
\usepackage{amsfonts}
\usepackage{graphicx}

\newcommand{\nt}{\notag}
\newcommand{\Ld}[1]{\frac{\overrightarrow{\delta}}{\delta #1}}
\newcommand{\Rd}[1]{\frac{\overleftarrow{\delta}}{\delta #1}}
\newcommand{\Tr}{\mathrm{Tr}\,}
\newcommand{\fs}[1]{{\hbox{$#1$\kern-0.4em\raise0.3ex\hbox{/}}}}
\newcommand{\vev}[1]{\left\langle #1 \right\rangle}
\newcommand{\vvev}[1]{\vev{\kern-0.3em\vev{#1}\kern-0.3em}}
\newcommand{\E}{\mathcal{E}}

\newcommand{\Op}{\mathcal{O}}
\newcommand{\op}[1]{\left[ #1 \right]}
\newcommand{\PP}{\mathcal{P}}
\newcommand{\lb}{\left\lbrace}
\newcommand{\rb}{\right\rbrace}
\newcommand{\ep}{\epsilon}

\begin{document}

\preprint{KOBE-TH-20-04}

\title{Products of Current Operators\\in the Exact Renormalization
  Group Formalism}


\author{H.~Sonoda}
\email[]{hsonoda@kobe-u.ac.jp}
\affiliation{Physics Department, Kobe University, Kobe 657-8501, Japan}


\date{\today}

\begin{abstract}
  Given a Wilson action invariant under global chiral transformations,
  we can construct current composite operators in terms of the Wilson
  action.  The short distance singularities in the multiple products
  of the current operators are taken care of by the exact
  renormalization group.  The Ward-Takahashi identity is compatible
  with the finite momentum cutoff of the Wilson action.  The exact
  renormalization group and the Ward-Takahashi identity together
  determine the products.  As a concrete
  example, we study the Gaussian fixed-point Wilson action of the
  chiral fermions to construct the products of current operators.
\end{abstract}


\maketitle


\newpage

\section{Introduction\label{sec-introduction}}

It is a principle of quantum field theory that the invariance of a
theory under a continuous transformation implies the conservation of a
current.  When a theory is expressed by a Wilson action with a finite
momentum cutoff, the principle holds for the Wilson action.  In
\cite{Sonoda:2015pva} an energy-momentum tensor was constructed from
the invariance of the Wilson action under translations and rotations.
In this paper we would like to consider the Wilson action of chiral
fermions with global flavor symmetry to construct multiple products of
the conserved current operator.

To build the Wilson action, we use the exact renormalization group
(ERG) formalism.  (See for example
\cite{Becchi:1996an,Berges:2000ew,Igarashi:2009tj,Rosten:2010vm} and  
references therein.)  The Wilson action satisfies a well defined
differential equation under the continuous change of scale.  We adopt
a convention that each time we integrate more of the
high-momentum fluctuations, we introduce a change of scale to restore
the same cutoff function.  The continuum limit corresponds to a
trajectory parametrized by a logarithmic scale parameter $t$ so that a
fixed point is reached in the limit $t \to - \infty$.

The Wilson action of a theory in the continuum limit has all the short
distance physics incorporated into the vertices of the action.  The
full theory is obtained by further integration of the fields with
momenta below the cutoff.  The Wilson action is determined by the ERG
differential equation whose solution is parametrized by the relevant
variables of the theory.

Composite operators can be considered as infinitesimal changes of the
Wilson action, and they also obey well defined differential equations
under the change of logarithmic scale.  The general properties of
products of composite operators have been discussed in
\cite{Pagani:2017tdr}.  We follow and extend the discussions there by
considering the multiple products of current operators.

ERG is good at handling the short distance singularities via ERG
differential equations.  Well defined ERG differential equations admit
only the solutions consistent with locality, i.e., the vertices of the
action and composite operators must be analytic at zero momenta.  This
is the guiding principle we follow throughout the paper.

Though we consider only chiral fermion fields as dynamical fields, our
discussion of current operators is easy to modify in the presence of other
dynamical fields, for example in the case of QCD with massless quarks.

Our subject obviously overlaps with the construction of chiral gauge
theories using the ERG formalism.  (See, for example,
\cite{Igarashi:2019gkm} and references therein.)  For example, the derivation of chiral
anomaly using the ERG formalism was done in the context of gauge
theory.  (See, for example, Sec.~9 of \cite{Igarashi:2009tj} and
\cite{Igarashi:2011xs}.)  The multiple products of current operators
require a much lighter formalism.

The paper is organized as follows.  In Sec.~II, we introduce a current
operator for a generic Wilson action of chiral fermions under the
assumption of global continuous symmetry.  In Sec.~III, we introduce
multiple products of current operators and derive the ERG equations
satisfied by them.  By coupling the current with an external gauge
field, we construct a composite operator in terms of which we can
consider all the products of current operators at once.  In
Sec.~IV, we consider the Ward-Takahashi (WT) identity satisfied by the
products of currents.  Our discussion is not completely closed, and we
need to introduce the identity as a working hypothesis.  The ERG
differential equation and the WT identity are mutually consistent, and
they together characterize the products of current operators.  In
Sec.~V we discuss the changes to the ERG equation and the WT identity
caused by the short-distance singularities of the operator products.
In Sec.~VI we consider the products of current operators for the free
theory.  Though this section is all about 1-loop diagrams, the example
elucidates the general formalism given in the preceding sections.

Please note that we use the following condensed notation for momentum
integrals:
\begin{equation}
  \int_p \equiv \int \frac{d^D p}{(2 \pi)^D},\quad
  \delta (p) \equiv (2\pi)^D \delta^{(D)} (p)\,.
\end{equation}

\section{Current composite operators\label{sec-current}}

We consider a theory of chiral fermion fields $\psi, \bar{\psi}$
satisfying
\begin{equation}
  a_R \psi (p) = \psi (p),\quad \bar{\psi} (-p) a_L = \bar{\psi} (-p), 
\end{equation}
where
\begin{equation}
  a_R \equiv \frac{1 + \gamma_5}{2},\quad a_L \equiv \frac{1 -
    \gamma_5}{2}.
\end{equation}
The theory is determined by its Wilson action with a fixed UV cutoff.
The cutoff is given in terms of a smooth momentum cutoff
function $K(p)$, such as $e^{- p^2}$, that is $1$ at $p=0$ and
vanishes as $p \to \infty$.  We parametrize the Wilson action by a
logarithmic scale parameter $t$ and demand that it obey the ERG
differential equation
\begin{align}
  \partial_t e^{S_t [\psi, \bar{\psi}]}
  &= \int_p \left[ \left( \frac{\Delta (p)}{K(p)} + \frac{D+1}{2} + p \cdot \partial_p -
    \gamma_t \right) \bar{\psi} (-p) \Ld{\bar{\psi} (-p)} e^{S_t}\right.\nt \\
  &  + e^{S_t} \Rd{\psi (p)} \left( \frac{\Delta (p)}{K(p)} +
    \frac{D+1}{2} + p \cdot \partial_p   - \gamma_t \right) \psi
    (p)\nt\\
  &\left. - \Tr \Ld{\bar{\psi} (-p)} e^{S_t} \Rd{\psi (p)} a_R
    \frac{\Delta (p) - 2 \gamma_t K(p) \left(1 - K(p)\right)}{\fs{p}} \right]\,,
\end{align}
where
\begin{equation}
  \Delta (p) \equiv - p \cdot \partial_p K(p)\,.
\end{equation}
$\gamma_t$ is an anomalous dimension of the chiral fermion field, the
trace is for both spinor and flavor indices, and the minus in front of
the trace is due to the Fermi statistics.

We assume that the Wilson action $S_t$ describes a continuum limit; as
we take $t \to - \infty$, we obtain a UV fixed-point:
\begin{equation}
  \lim_{t \to - \infty} S_t = S^*\,.
\end{equation}
All the physics beyond the fixed cutoff scale of $1$ has been
incorporated into the action.  By integrating the fluctuations of
momenta less than $1$, we get full correlation functions of the
fields.

We define the correlation functions by
\begin{align}
&  \vvev{\psi (p_1) \cdots \psi (p_n) \bar{\psi} (-q_1) \cdots
    \bar{\psi} (-q_n)}_t\equiv \prod_{i=1}^n \frac{1}{K (p_i) K(q_i)}
                \nt\\
  &\times  \vev{\psi
        (p_1) \cdots \psi (p_n)  \exp \left( - \int_p \Rd{\psi (p)}
          K(p) h_F (p)  \Ld{\bar{\psi} (-p)}
          \right) \bar{\psi} (-q_1) \cdots 
        \bar{\psi} (-q_n)}_{S_t}\,,
\end{align}
where
\begin{equation}
  h_F (p) \equiv a_R \frac{1-K(p)}{\fs{p}}
\end{equation}
is the high-momentum propagator.  The correction involving the cutoff
function is a technicality typical in the ERG formalism.  Thanks to
the correction, though, the correlation functions satisfy the simple
scaling relation
\begin{align}
  & \vvev{\psi (p_1 e^{t-t'}) \cdots \bar{\psi} (-q_n e^{t-t'})}_t\nt\\
  &= \exp \left( - n (D-1) (t-t') + 2n \int_{t'}^t d\tau \, \gamma_\tau \right)
    \vvev{\psi (p_1) \cdots \bar{\psi} (-q_n)}_{t'}\,.
\end{align}

Another technicality is necessary before we move on to discuss
symmetry.  A composite operator $\Op_t (p)$ is a functional whose
correlation functions are defined by
\begin{align}
  &\vvev{\Op_t (p)\,\psi (p_1) \cdots \bar{\psi} (-q_n)}_t
    \equiv \prod_{i=1}^n \frac{1}{K(p_i) K(q_i)}\nt\\
  &\equiv \vev{ \Op_t (p)\, \left( \psi (p_1) \cdots \exp \left( -
    \int_q \Rd{\psi (q)} K(q) h_F (q) \Ld{\bar{\psi} (-q)} \right)
    \cdots \bar{\psi} (-q_n) \right)}_{S_t}
\end{align}
where the exponentiated differential operator does not act on
$\Op_t (p)$.  We define $\Op_t (p)$ so that its correlation functions
satisfy the scaling relation
\begin{align}
&  \vvev{\Op_t (p e^{t-t'}) \psi (p_1 e^{t-t'}) \cdots \bar{\psi} (-q_n
    e^{t-t'})}_t\nt\\
&  = e^{- y (t-t')} \exp \left( - n (D-1) (t-t') + n \int_{t'}^t
    d\tau\,\gamma_\tau \right) \vvev{\Op_{t'} (p) \psi (p_1) \cdots
    \bar{\psi} (-q_n)}_{t'} \label{sec2-Op-scaling}
\end{align}
For simplicity we have taken $-y$, the scale dimension of $\Op_t (p)$,
independent of $t$.  For (\ref{sec2-Op-scaling}) to be valid, 
$\Op_t (p)$ must satisfy the ERG differential equation
\begin{equation}
  \left(\partial_t + y + p \cdot \partial_p - \mathcal{D}_t \right) \Op_t (p)
  = 0\,,
\end{equation}
where $\mathcal{D}_t$, acting on functionals, is defined by
\begin{align}
  \mathcal{D}_t \Op &\equiv
                \int_q \left[ \left( \frac{\Delta (q)}{K(q)} +
                \frac{D+1}{2}  - \gamma_t + q \cdot \partial_q \right)
                \bar{\psi} (-q) \cdot \Ld{\bar{\psi} (-q)} \Op \right.\nt\\
  &\qquad + \Op \Rd{\psi (q)} \left( \frac{\Delta (q)}{K(q)} +
                \frac{D+1}{2}  - \gamma_t + q \cdot \partial_q \right)
    \psi (q)\nt\\
  &\qquad+ S_t \Rd{\psi (q)}
    \left(a_R \frac{\Delta (q)}{\fs{q}} - 2 \gamma_t K(q) h_F (q) \right)
    \Ld{\bar{\psi} (-q)} \Op\nt \\
   &\qquad + \Op \Rd{\psi (q)} \left( a_R \frac{\Delta (q)}{\fs{q}} - 2 \gamma_t
    K(q) h_F (q)\right) \Ld{\bar{\psi} (-q)} S_t \nt\\
  &\left.\qquad - \Tr \left( a_R\frac{\Delta (q)}{\fs{q}} - 2 \gamma_t K(q) h_F
    (q) \right) \Ld{\bar{\psi} (-q)} \Op \Rd{\psi (q)} \right]\,.\label{sec2-Dt}
\end{align}
The simplest example of a composite operator is
\begin{subequations}
  \label{sec2-PsiPsibar}
\begin{align}
  \Psi (p)
  &\equiv \frac{1}{K(p)} \left[ \psi (p) + h_F (p) \Ld{\bar{\psi}
    (-p)} S_t\right]\,,\\
  \bar{\Psi} (-p)
  &\equiv \frac{1}{K(p)} \left[ \bar{\psi} (-p) + S_t \Rd{\psi (p)}
    h_F (p)\right] \,.
\end{align}
\end{subequations}
Though they are composite operators, they have the same correlation
functions as the elementary fields $\psi (p), \bar{\psi} (-p)$:
\begin{align}
  \vvev{\Psi (p_1) \psi (p_2) \cdots \bar{\psi} (-q_n)}_t
  &= \vvev{\psi (p_1) \cdots \bar{\psi} (-q_n)}_t\,,\\
  \vvev{\psi (p_1) \cdots \bar{\psi} (-q_{n-1}) \bar{\Psi} (-q_n)}_t
  &= \vvev{\psi (p_1) \cdots \bar{\psi} (-q_n)}_t\,.
\end{align}

We are now ready to discuss symmetry.  We assume that the correlation
functions have global symmetry:
\begin{equation}
  \vvev{U \psi (p_1) \cdots U \psi (p_n) \,\bar{\psi} (-q_1) U^\dagger
    \cdots \bar{\psi} (-q_n) U^\dagger}_t
  = \vvev{\psi (p_1) \cdots \psi (p_n) \bar{\psi} (-q_1) \cdots
    \bar{\psi} (-q_n)}_t\,,\label{sec2-symmetry}
\end{equation}
where $U$ is an arbitrary unitary matrix that acts on the flavor
indices of $\psi$ and $\bar{\psi}$.  ($U$ may be a U($N$) matrix if we
have $N$ flavors.)  For infinitesimal transformations we obtain
\begin{align}
  &\sum_{i=1}^n \left(
  - \vvev{\psi (p_1) \cdots T^a \psi (p_i) \cdots \psi (p_n)
  \bar{\psi} (-q_1) \cdots \bar{\psi} (-q_n)}_t \right.\nt\\
  &\quad \left. + \vvev{\psi (p_1) \cdots \psi (p_n) \bar{\psi} (-q_1)
    \cdots \bar{\psi} (-q_i) T^a \cdots \bar{\psi} (-q_n)}_t \right) = 0\,,
\end{align}
where $T^a$ are hermitian matrices normalized by
\begin{equation}
  \Tr T^a T^b = \delta^{ab}
\end{equation}
and satisfying the commutation relation
\begin{equation}
  \op{ T^a , T^b } = i \sum_c f^{abc} T^c\,.
\end{equation}
(We will omit the summation symbol for the repeated indices $c$ from now on.)

To express (\ref{sec2-symmetry}) as an operator equation, 
we introduce an equation-of-motion composite operator by
\begin{equation}
  \E^a (p) \equiv e^{-S_t} \int_q K(q) \Tr \left[
    \Ld{\bar{\psi} (-q)} \left( \bar{\Psi} (-q+p) T^a e^{S_t}
    \right)
    - \left( e^{S_t} T^a \Psi (q+p) \right) \Rd{\psi (q)} \right]\,,
\end{equation}
where $\Psi, \bar{\Psi}$ are defined by (\ref{sec2-PsiPsibar}).
$\E^a$ is a total derivative of the
exponentiated Wilson action, and it has correlation functions
\begin{align}
&  \vvev{\E^a (p)\, \psi (p_1) \cdots \psi (p_n) \bar{\psi} (-q_1)
  \cdots \bar{\psi} (-q_n)}_t\nt\\
  &= \sum_{i=1}^n \left[ - \vvev{\psi (p_1) \cdots T^a \psi (p+p_i)
    \cdots \psi (p_n) \bar{\psi} (-q_1) \cdots \bar{\psi} (-q_n)}_t\right.
    \nt\\
  &\left.\quad + \vvev{\psi (p_1) \cdots \psi (p_n) \bar{\psi} (-q_1)
    \cdots \bar{\psi} (p-q_i) T^a \cdots \bar{\psi} (-q_n)}_t\right]\,.
\end{align}
The symmetry (\ref{sec2-symmetry}) is equivalent to
\begin{equation}
  \E^a (p=0) = 0\,.\label{sec2-symmetry-Wilson}
\end{equation}
In fact this is equivalent to what we usually consider as the invariance of the action
\begin{equation}
  \int_p \left( \bar{\psi} (-p) T^a \Ld{\bar{\psi} (-p)} S_t - S_t
    \Rd{\psi (p)} T^a \psi (p) \right) = 0\,.
\end{equation}
In Appendix \ref{appendix-invariance-action} we show that this is
equivalent to (\ref{sec2-symmetry-Wilson}) .

Since $\E^a (p)$ is a local operator, it must be proportional to the
momentum:
\begin{equation}
  \E^a (p) = p_\mu J_\mu^a (p)\,,\label{sec2-pmuJmu}
\end{equation}
where the current $J_\mu^a (p)$ must be a local composite operator.
Unless there is a local operator $j_\mu^a (p)$ orthogonal to $p_\mu$
\begin{equation}
  p_\mu j_\mu^a (p) = 0\,,
\end{equation}
(\ref{sec2-pmuJmu}) defines the current $J_\mu^a (p)$ unambiguously.
Since $\E^a (p)$ has scale dimension $0$, $J_\mu^a (p)$ must have
scale dimension $-1$.  In coordinate space
$J_\mu^a (x) = \int_p e^{i p x} J_\mu^a (p)$ has scale dimension
$D-1$.

As an example, let us consider the Gaussian fixed-point theory
\begin{equation}
  S_G = - \int_p \frac{1}{K(p)} \bar{\psi} (-p) \fs{p} a_R \psi (p)\,,
\end{equation}
for which
\begin{equation}
  \Psi (p) = \psi (p)\,,\quad \bar{\Psi} (-p) = \bar{\psi} (-p)\,.
\end{equation}
We find
\begin{align}
  \E^a (p) &=  \int_q K(q) \left[ - \bar{\psi} (-q+p) T^a \Ld{\bar{\psi}
             (-q)} S + S \Rd{\psi (q)} T^a \psi (q+p) \right]\nt\\
           &= \int_q \left( \bar{\psi} (-q+p) T^a a_R \fs{q} \psi (q) -
             \bar{\psi} (-q) a_R \fs{q} T^a \psi (q+p) \right)\nt\\
  &= \int_q \bar{\psi} (-q) T^a a_R \fs{p} \psi (q+p) \,.
\end{align}
This implies
\begin{equation}
J_\mu^a (p) = \int_q \bar{\psi} (-q) T^a \gamma_\mu a_R \psi
(q+p)\,.\label{sec2-current-Gauss} 
\end{equation}

\section{Products of current operators}

We wish to define multiple products of currents.  The
product of two currents is defined as
\begin{equation}
  \op{J_\mu^a (p) J_\nu^b (q)} \equiv J_\mu^a (p) J_\nu^b (q) + 
  \PP_{\mu\nu}^{ab} (p,q) \,,
\end{equation}
where $\PP$ is a local counterterm necessary to make the product a
composite operator; the bare product $J_\mu^a (p) J_\nu^b (q)$ is not
a composite operator in the sense introduced in the previous section.
$\PP$ also takes care of the short-distance singularity occurring when
the two currents come close together.  For the product to be a
composite operator of scale dimension $-2$, it must satisfy
\begin{equation}
  \left( \partial_t + p \cdot \partial_p + q \cdot \partial_q + 2 -
    \mathcal{D}_t \right) \op{J_\mu^a (p) J_\nu^b (q)} = 0\,,
\end{equation}
where $\mathcal{D}_t$ is given by (\ref{sec2-Dt}).  This implies
\begin{align}
&   \left( \partial_t + p \cdot \partial_p + q \cdot \partial_q + 2 -
                \mathcal{D}_t \right) \PP_{\mu\nu}^{ab} (p,q) \nt\\
  &=  \int_r \left[ J_\mu^a (p) \Rd{\psi (r)} \left( a_R \frac{\Delta
       (r)}{\fs{r}} - 2 \gamma_t K(r) h_F (r) \right) \Ld{\bar{\psi}
     (-r)} J_\nu^b (q) + \left( J_\mu^a (p) \leftrightarrow J_\nu^b
    (q)\right) \right]\,.
\end{align}

Similarly, we define the product of three currents as
\begin{align}
  &\op{J_{\mu_1}^{a_1} (p_1) J_{\mu_2}^{a_2} (p_2) J_{\mu_3}^{a_3}
  (p_3)}
  \equiv J_{\mu_1}^{a_1} (p_1) J_{\mu_2}^{a_2} (p_2) J_{\mu_3}^{a_3}
    (p_3)\nt\\
  &\quad + \PP_{\mu_1\mu_2}^{a_1 a_2} (p_1,p_2) J_{\mu_3}^{a_3} (p_3)
    + \PP_{\mu_2\mu_3}^{a_2 a_3} (p_2,p_3) J_{\mu_1}^{a_1} (p_1)
    + \PP_{\mu_3 \mu_1}^{a_3 a_1} (p_3, p_1) J_{\mu_2}^{a_2} (p_2)\nt\\
  &\quad + \PP_{\mu_1\mu_2\mu_3}^{a_1a_2a_3} (p_1,p_2,p_3)\,,
\end{align}
and so on for the higher order products.  We note that
$\PP_{\mu_1\cdots\mu_n}^{a_1 \cdots a_n} (p_1, \cdots, p_n)$
gives the short-distance singularity due to all the $n$ currents
coming together simultaneously, and it is proportional to the delta function in
momentum space
\begin{equation}
  \PP_{\mu_1\cdots\mu_n}^{a_1 \cdots a_n} (p_1, \cdots, p_n) \propto
  \delta \left(\sum_{i=1}^n p_i \right)\,,
\end{equation}
unless there is a composite operator of scale dimension $-n$ or less
available.  (That means scale dimension $D-n$ or less in coordinate
space.)  The ERG equation for
$\PP_{\mu_1 \mu_2 \mu_3} (p_1, p_2, p_3)$ is given by
\begin{align}
  & \left( \partial_t + \sum_{i=1}^3 p_i \cdot \partial_{p_i} + 3 -
    \mathcal{D}_t \right) \PP_{\mu_1 \mu_2 \mu_3}^{a_1 a_2 a_3} (p_1,
    p_2, p_3)\nt\\
  &= \int_q \Big[ \PP_{\mu_1 \mu_2}^{a_1 a_2} (p_1, p_2) \Rd{\psi
    (q)} \left( a_R \frac{\Delta (q)}{\fs{q}} - 2 \gamma_t K(q) h_F
    (q) \right) \Ld{\bar{\psi} (-q)} J_{\mu_3}^{a_3} (p_3) \nt\\
  &\quad + J_{\mu_3}^{a_3} (p_3) \Rd{\psi
    (q)} \left( a_R \frac{\Delta (q)}{\fs{q}} - 2 \gamma_t K(q) h_F
    (q) \right) \Ld{\bar{\psi} (-q)}  \PP_{\mu_1 \mu_2}^{a_1 a_2}
    (p_1, p_2) \nt\\
  &\quad + (\textrm{4 other terms}) \Big]\,.
\end{align}
The ERG equations for the higher order counterterms are given similarly.

To consider all the local products of current operators
simultaneously, we introduce a classical gauge field coupled to the
current:
\begin{equation}
 W_t [A_\mu^a] \equiv \int_p A_\mu^a (-p) J_\mu^a (p)
+ \sum_{n=2}^\infty \frac{1}{n!}
  \int_{p_1,\cdots,p_n} A_{\mu_1}^{a_1} (-p_1) \cdots A_{\mu_n}^{a_n}
  (-p_n) \, \PP_{\mu_1\cdots\mu_n}^{a_1\cdots a_n} (p_1,\cdots,p_n) 
\end{equation}
so that its exponential
\begin{equation}
 e^{W_t [A]} \equiv \sum_{n=0}^\infty 
 \frac{1}{n!}
  \int_{p_1,\cdots,p_n} A_{\mu_1}^{a_1} (-p_1) \cdots A_{\mu_n}^{a_n}
  (-p_n) \, \op{J_{\mu_1}^{a_1} (p_1) \cdots J_{\mu_n}^{a_n} (p_n)} 
\end{equation}
is a composite operator.  We assign the scale dimension $-D+1$ to the
source field $A_\mu^a$ so that $e^{W_t [A]}$ becomes a composite operator
of scale dimension $0$, satisfying the ERG equation
\begin{equation}
  \left( \partial_t + \int_p \left( - p \cdot \partial_p - D + 1 \right)
  A_\mu^a (p) \cdot \frac{\delta}{\delta A_\mu^a (p)}  - \mathcal{D}_t
\right)
e^{W_t [A]} = 0\,,\label{sec3-Wt-ERG}
\end{equation}
where $\mathcal{D}_t$ is defined by (\ref{sec2-Dt}).

\section{Commutation relation --- Ward-Takahashi identity}

We now wish to consider the ``commutation relation'' of two currents.
The quotation mark is put because it needs to be explained.  Our
commutation relation is an operator equation
\begin{equation}
  p_\mu \op{ J_\mu^a (p) J_\nu^b (q)} = i f^{abc} J_\nu^c (p+q) + \E^a
  (p) \star J_\nu^b (q)\label{sec4-WTJJ}
\end{equation}
which amounts to the Ward-Takahashi (WT) identity
\begin{align}
&  p_\mu \vvev{\op{J_\mu^a (p) J_\nu^b (q)} \psi (p_1) \cdots
  \bar{\psi} (-q_n)}_t = i f^{abc} \vvev{J_\nu^c (p+q) \psi (p_1)
                \cdots \bar{\psi} (-q_n)}_t\nt\\
  &\quad + \sum_{i=1}^n \left( - \vvev{J_\nu^b (q) \psi (p_1) \cdots
    T^a \psi (p+p_i) \cdots \bar{\psi} (-q_n)}_t\right.\nt\\
  &\qquad\qquad \left. + \vvev{J_\nu^b (q) \psi (p_1) \cdots \bar{\psi}
    (p-q_i) T^a \cdots \bar{\psi} (-q_n)}_t\right) \,.\label{sec4-Ward}
\end{align}
We wish to explain the above and its generalization to higher order
products in this section.

We define an equation-of-motion composite operator by
\begin{align}
  \E^a (p) \star J_\alpha^b (q)
  &\equiv e^{-S} \int_r K(r) \Tr \left[ \Ld{\bar{\psi} (-r)} \left(
    \op{\bar{\Psi} (-r+p) T^a J_\alpha^b (q)} e^{S_t} \right)\right.\nt\\
  &\left.\qquad\qquad  - \left( e^{S_t} \op{T^a \Psi (q+p) J_\alpha^b (q)}
    \right) \Rd{\psi (r)} \right]\,,
\end{align}
where
\begin{align}
  \op{\bar{\Psi} (-r) J_\alpha^b (q)}
  &\equiv \bar{\Psi} (-r) J_\alpha^b (q) + J_\alpha^b (q) \Rd{\psi (r)} h_F (r)\,,\\
  \op{\Psi (r) J_\alpha^b (q)}
  &\equiv \Psi (r) J_\alpha^b (q) + h_F (r) \Ld{\bar{\psi} (-r)}
    J_\alpha^b (q)
\end{align}
are the composite operators satisfying
\begin{align}
  \vvev{\psi (p_1) \cdots
  \bar{\psi} (-q_{n-1}) \op{\bar{\Psi} (-r) J_\alpha^b (q)}}_t
  &= \vvev{J_\alpha^b (q) \psi (p_1) \cdots \bar{\psi} (-q_{n-1})
    \bar{\psi} (-r)}_t\,,\\
  \vvev{\op{\Psi (r) J_\alpha^b (q)} \psi (p_2) \cdots
  \bar{\psi} (-q_{n})}_t
  &=  \vvev{J_\alpha^b (q) \psi (r) \psi (p_2) \cdots \bar{\psi} (-q_n)}_t\,.
\end{align}
Hence, we obtain
\begin{align}
&  \vvev{\E^a (p) \star J_\alpha^b (q)\, \psi (p_1) \cdots \bar{\psi}
                (-q_n)}_t\nt\\
  &=  \sum_{i=1}^n \left( - \vvev{J_\nu^b (q) \psi (p_1) \cdots
    T^a \psi (p+p_i) \cdots \bar{\psi} (-q_n)}_t\right.\nt\\
  &\qquad\qquad \left. + \vvev{J_\nu^b (q) \psi (p_1) \cdots \bar{\psi}
    (p-q_i) T^a \cdots \bar{\psi} (-q_n)}_t\right) \,.
\end{align}
This gives the second term of the right-hand side of
(\ref{sec4-Ward}).

Let
\begin{equation}
  \Op_\nu^{ab} (p+q) \equiv p_\mu \op{J_\mu^a (p) J_\nu^b (q)} - \E^a (p)
  \star J_\nu^b (q)\,.  \label{sec4-Oab}
\end{equation}
(\ref{sec4-Ward}) then amounts to
\begin{equation}
  \Op^{ab}_\nu (p+q) = i f^{abc} J_\nu^c (p+q)\,.\label{sec4-equality}
\end{equation}
This equality is plausible but not obvious, and it needs an
explanation.  We will check this later explicitly for the Gaussian
theory.  Here we satisfy ourselves by checking the consistency of
(\ref{sec4-equality}) with Bose symmetry of the current operator,
which requires the product
\[
p_\mu q_\nu \op{J_\mu^a (p) J_\nu^b (q)}
\]
be symmetric under the interchange.  The product may depend on which
divergence we calculate first.  Calculating $p_\mu J_\mu^a (p)$ first,
(\ref{sec4-Oab}) gives
\begin{align}
  p_\mu q_\nu \op{J_\mu^a (p) J_\nu^b (q)}
  &= q_\nu \Op_\nu^{ab} (p+q) + \E^a (p) \star q_\nu J_\nu^b (q) \nt\\
  &= q_\nu \Op_\nu^{ab} (p+q) + \E^a (p) \star \E^b (q)\,.
\end{align}
Calculating $q_\nu J_\nu^b (q)$ first, we obtain
\begin{align}
  p_\mu q_\nu \op{J_\mu^a (p) J_\nu^b (q)}
  &= p_\mu \Op_\mu^{ba} (p+q) + \E^b (q) \star p_\mu J_\mu^a (p)\nt\\
  &= p_\mu \Op_\mu^{ba} (p+q) + \E^b (q) \star \E^a (p)\,.
\end{align}
Hence, for consistency, we must find
\begin{equation}
  p_\mu \Op_\mu^{ba} (p+q) - q_\mu \Op_\mu^{ab} (p+q) = \E^a (p) \star
  \E^b (q) - \E^b (q) \star \E^a (p)\,.\label{sec4-consistency}
\end{equation}
To compute the right-hand side, we consider correlation functions:
\begin{align}
&  \vvev{\E^a (p) \star \E^b (q)\,\psi (p_1) \cdots \psi (p_n)
                \bar{\psi} (-q_1) \cdots \bar{\psi} (-q_n)}_t\nt\\
  &= \sum_{i=1}^n \left[ - \vvev{\E^b (q)\, \cdots T^a \psi (p+p_i)
    \cdots}_t + \vvev{\E^b (q)\, \cdots \bar{\psi} (p-q_i) T^a
    \cdots}_t \right]\nt\\
  &= \sum_{i=1}^n \Bigg[ \vvev{\cdots T^a T^b \psi (p+q+p_i) \cdots}_t
    + \vvev{\cdots \bar{\psi} (p+q-q_i) T^b T^a \cdots}_t\nt\\
  &\quad - \sum_{j=1}^n \left( \vvev{\cdots T^a \psi (p+p_i) \cdots
    \bar{\psi} (q-q_i) T^b \cdots}_t + \vvev{\cdots T^b \psi (q+p_i)
    \cdots \bar{\psi} (p-q_i) T^b \cdots}_t\right)\nt\\
  &\quad + \sum_{j\ne i} \left( \vvev{\cdots T^a \psi (p+p_i) \cdots
    T^b \psi (q+p_j) \cdots}_t + \vvev{\cdots \bar{\psi} (p-q_i) T^a
    \cdots \bar{\psi} (q-q_i) T^b \cdots}_t \right) \Bigg]
    \,.
\end{align}
Hence, we obtain
\begin{equation}
  \E^a (p) \star \E^b (q) - \E^b (q) \star \E^a (p) = - i f^{abc} \E^c
  (p+q)\,.\label{sec4-EE-commutator}
\end{equation}
Then, the consistency condition (\ref{sec4-consistency}) gives
\begin{equation}
p_\mu \Op_\mu^{ba} (p+q) - q_\mu \Op_\mu^{ab} (p+q) = - i f^{abc} \E^c
(p+q)
=(- i f^{abc})  (p+q)_\mu  J_\mu^c (p+q)\,,
\end{equation}
which is indeed satisfied by (\ref{sec4-equality}).

We have thus checked at least that (\ref{sec4-WTJJ}) is consistent
with the Bose symmetry of the current.  We adopt (\ref{sec4-WTJJ})
and its generalization to higher orders as our working hypothesis:
\begin{align}
  p_\mu \op{J_\mu^a (p) J_{\mu_1}^{a_1} (p_1) \cdots J_{\mu_k}^{a_k}
    (p_k)}
  &= \sum_{i=1}^k i f^{a a_i b} \op{J_{\mu_1}^{a_1} (p_1) \cdots
    J_{\mu_i}^b (p+p_i) \cdots J_{\mu_k}^{a_k} (p_k)}\nt\\
&\quad  + \E^a (p) \star \op{J_{\mu_1}^{a_1} (p_1) \cdots
                                                             J_{\mu_k}^{a_k} (p_k)}\,.
                                                             \label{sec4-WT}
\end{align}
For the correlation functions, this gives
\begin{align}
&  p_\mu \vvev{\op{J_\mu^a (p) J_{\mu_1}^{a_1} (p_1) \cdots J_{\mu_k}^{a_k}
                 (p_k)}\psi (q_1) \cdots \psi (q_n) \bar{\psi}
                 (-r_1) \cdots \bar{\psi} (-r_n)}_t\nt\\
  &= \sum_{i=1}^k i f^{a a_i b} \vvev{\op{J_{\mu_1}^{a_1} (p_1) \cdots
    J_{\mu_i}^b (p_i+p) \cdots J_{\mu_k}^{a_k} (p_k)} \psi (q_1) \cdots \psi (q_n) \bar{\psi}
    (-r_1) \cdots \bar{\psi} (-r_n)}_t\nt\\
  &\quad + \sum_{j=1}^n \lb - \vvev{\op{J_{\mu_1}^{a_1} (p_1) \cdots
    J_{\mu_k}^{a_k} (p_k)} \cdots T^a \psi (q_j+p) \cdots}_t\right.\nt\\
&\qquad\qquad\left. + \vvev{\op{J_{\mu_1}^{a_1} (p_1) \cdots
    J_{\mu_k}^{a_k} (p_k)} \cdots \bar{\psi} (-r_j+p) T^a \cdots}_t \rb\,.
\end{align}

The WT identity (\ref{sec4-WT}) we just introduced is compactly
expressed in terms of the composite operator $e^{W_t [A]}$ as
\begin{equation}
  \int_q \left( p_\mu \delta_{ab} \delta (p-q) + i f^{acb} A_\mu^c (-q+p)
  \right) \frac{\delta}{\delta A_\mu^b (-q)} e^{W_t [A]} 
  = \E^a (p) \star e^{W_t [A]}  \,.
\end{equation}
Expanding this in powers of the external source $A$, we can easily
check the equivalence to (\ref{sec4-WT}).  Multiplying an infinitesimal
$\ep^a (-p)$ and integrating over $p$, we can rewrite this as
\begin{equation}
 \delta_\ep e^{W_t [A]} \equiv  e^{W_t \left[ A^\ep \right]} - e^{W_t [A]} = \int_p \ep^a (-p) \E^a
  (p) \star e^{W_t [A]}\,,\label{sec4-Wt-WT}
\end{equation}
where
\begin{equation}
  \left(A^\ep\right)_\mu^a (-p) \equiv A_\mu^a (-p) + p_\mu \ep^a (-p) +
  i f^{abc} \int_q   A_\mu^b (q-p) \ep^c (-q)
\end{equation}
is an infinitesimal gauge transformation.

\section{Corrections to the ERG equation and the WT identity}

We have identified two important properties of $e^{W_t [A]}$.  One is
the ERG differential equation (\ref{sec3-Wt-ERG}), and the other is
the gauge invariance (\ref{sec4-Wt-WT}).  Both may receive corrections
due to short distance singularities.  Since the nature of
singularities depends on the space dimension $D$, we specify $D=4$ in
the following discussion.

We first consider possible corrections to the ERG equation.  The
product of $n$ current operators has scale dimension $-n$, and it can
mix with operators of the same scale dimension.  As for the mixing
with the delta function $\delta (\sum_i p_i)$, we only need to
consider
\[
  \op{J_{\alpha}^a (p_1) J_\beta^b (p_2)},\quad
  \op{J_\alpha^a (p_1) J_\beta^b (p_2) J_\gamma^c (p_3)},\quad
  \op{J_\alpha^a (p_1) J_\beta^b (p_2) J_\gamma^c (p_3) J_\delta^d
    (p_4)}
\]
which mix with the delta function $\delta (\sum_i p_i)$ with
appropriate powers (quadratic, linear, none) of momenta.  This gives a
new ERG differential equation:
\begin{equation}
  \left( \partial_t + \int_p \left( - p \cdot \partial_p - D + 1
    \right) A_\mu^a (p) \cdot \frac{\delta}{\delta A_\mu^a (p)} -
    \mathcal{D}_t \right) e^{W_t [A]}
  = \int d^D x\, f (t; A (x))\, e^{W_t [A]}\,,\label{sec5-Wt-ERG}
\end{equation}
where $f$ is a linear combination of the products of two $A$'s with
two derivatives, three $A$'s with one derivative, and four $A$'s with
no derivative.  Consistency with (\ref{sec4-Wt-WT}) gives the gauge
invariance of $f$.  Hence, we obtain
\begin{equation}
 f (t; A) = b (t) \frac{1}{4} \int d^4 x\, \Tr \left( \partial_\alpha A_\beta -
   \partial_\beta A_\alpha - i [A_\alpha, A_\beta] \right)^2\,,\label{sec5-Wt-ERG-ft}
\end{equation}
where
\begin{equation}
  A_\mu \equiv T^a A_\mu^a\,.
\end{equation}

In fact the gauge invariance (\ref{sec4-Wt-WT}) itself may also get
corrected as
\begin{equation}
  \delta_\ep e^{W_t[A]} \equiv  e^{W_t[A^\ep]} - e^{W_t [A]} =
  \int_p \ep^a (-p) \left( \E^a (p) \star  + F^a (p; A) \right) e^{W_t[A]}\,,
  \label{sec5-Wt-WT}
\end{equation}
where $F^a (p; A)$ is a polynomial of $A$ with scale dimension $-4$.
This is the familiar chiral anomaly.  \cite{Adler:1969gk,Bell:1969ts}
For (\ref{sec5-Wt-WT}) to be consistent with (\ref{sec5-Wt-ERG}),
$F^a (p; A)$ must be independent of $t$, i.e., the anomaly must be
scale independent.

The algebraic structure of the anomaly is well
known.\cite{Wess:1971yu} For completeness, let us derive it using the
ERG formalism.  By definition of $\delta_\ep$, we must obtain
\begin{equation}
  \left( \delta_\eta \delta_\ep - \delta_\ep \delta_\eta \right)
  e^{W_t[A]} = \delta_{\op{\eta, \ep}} e^{W_t [A]}\,,\label{sec5-consistency}
\end{equation}
where
\begin{equation}
  \op{\eta, \ep} = \eta^a \ep^b \op{ T^a, T^b } = i f^{abc} \eta^a
  \ep^b T^c\,.
\end{equation}
Using (\ref{sec5-Wt-WT}) twice, we obtain
\begin{align}
  \left(\delta_\ep \delta_\eta - \delta_\eta \delta_\ep \right) e^{W_t
  [A]}
  &= \int_p \ep^a (-p) \int_q \eta^b (-q) (-i) f^{abc} \E^c (p+q)
    \star e^{W_t[A]}\nt\\
  &\quad + \int_p \left( \ep^a (-p) \delta_\eta F^a (p;A) - \eta^a
    (-p) \delta_\ep F^a (p;A) \right) e^{W_t[A]}\,,
\end{align}
where we have used (\ref{sec4-EE-commutator}).  Hence,
(\ref{sec5-consistency}) gives the desired algebraic constraint
\begin{align}
  \int_p \left( \ep^a (-p) \delta_\eta F^a (p;A) - \eta^a (-p)
  \delta_\ep F^a (p;A) \right)  = - i f^{abc} \int_p \ep^a (-p) \eta^b
  (-q) F^c (p+q; A)\,.
\end{align}
A well-known nontrivial solution to this is given by \cite{Bardeen:1969md}
\begin{equation}
  \int_p \ep^a (-p) F^a (p)
= \mathrm{const} \times \ep_{\alpha\beta\gamma\delta} \int d^4 x\,
    \Tr \partial_\alpha \ep \cdot \left( A_\beta \partial_\gamma
      A_\delta + \frac{1}{2 i} A_\beta A_\gamma A_\delta \right)\,.\label{sec5-anomaly}
\end{equation}  
(A trivial solution is $\delta_\ep$ of a polynomial of $A$.)

Concluding this section, we have explained that the ERG equation for
$e^{W_t [A]}$ can be modified to (\ref{sec5-Wt-ERG}) and
(\ref{sec5-Wt-ERG-ft}), and that the WT identity can get an anomaly
(\ref{sec5-Wt-WT}) where $F^a$ is given by (\ref{sec5-anomaly}).
Differentiating these with respect to the source $A$, we can get the
ERG equation and WT identity satisfied by the products of the current
operators.  Since their expressions are lengthy, we give them in
Appendix \ref{appendix-equations-for-J}.

\section{Free theory in $D=4$}

As a concrete example, we construct $W[A]$ for the Gaussian
fixed-point theory in $D=4$:
\begin{equation}
  W[A] = \int_p A_\mu^a (-p) J_\mu^a (p) + \sum_{n=2}^\infty
  \frac{1}{n!} \int_{p_1, \cdots, p_n} A_{\mu_1}^{a_1} (-p_1) \cdots
  A_{\mu_n}^{a_n} (-p_n)\, \PP_{\mu_1 \cdots \mu_n}^{a_1 \cdots a_n}
  (p_1, \cdots, p_n)\,.
\end{equation}
The construction of $e^{W[A]}$ is guided by two equations.  One is the
ERG differential equation
\begin{align}
&  \left( \int_p \left( - p \cdot \partial_p - D+1\right) A_\mu^a (p)
  \cdot \frac{\delta}{\delta A_\mu^a (p)} - \mathcal{D} \right)
  e^{W[A]}\nt\\
&\qquad = \frac{b}{4} \int d^4 x\, \Tr \left( \partial_\alpha A_\beta - \partial_\beta
    A_\alpha - i \op{A_\alpha, A_\beta} \right)^2\,,\label{sec6-W-ERG}
\end{align}
where $b$ is a constant, and $\mathcal{D}$ is defined by
\begin{align}
  \mathcal{D} \Op
  &\equiv \int_q \left[ \left(\frac{D+1}{2}
    + q \cdot \partial_q \right) \bar{\psi} (-q) \cdot \Ld{\bar{\psi}
    (-q)} \Op +  \Op \Rd{\psi (q)} \left(  \frac{D+1}{2} + q \cdot
    \partial_q \right) \psi (q)\right.\nt\\ 
&\left.\qquad - \Tr a_R \frac{\Delta (q)}{\fs{q}} \Ld{\bar{\psi} (-q)} \Op
          \Rd{\psi (q)}\right]\,.
\end{align}
The other is the WT identity with anomaly
\begin{equation}
  \delta_\ep e^{W[A]}
  = \left[ \int_p \ep^a (-p) \E^a (p) \star  + \mathcal{A}\,
  \ep_{\alpha\beta\gamma\delta} \int d^4 x\, \Tr \partial_\alpha \ep
  \cdot \left( A_\beta \partial_\gamma A_\delta + \frac{1}{2i} A_\beta
    A_\gamma A_\delta \right)\right] e^{W[A]} \,,\label{sec6-W-WT}
\end{equation}
where $\mathcal{A}$ is a constant.  Both $b$ and $\mathcal{A}$ are
determined as we construct
$\PP_{\mu_1 \cdots \mu_n}^{a_1 \cdots a_n} (p_1, \cdots, p_n)$ from
$n=2$ to higher $n$.  $b$ is determined by locality.  Locality implies
the analyticity of $\PP$'s at zero momenta.  We must choose $b$
appropriately to guarantee that (\ref{sec6-W-ERG}) admits a solution
satisfying locality.  Similarly, the coefficient $\mathcal{A}$ of the
chiral anomaly is determined by locality.  The solution to
(\ref{sec6-W-ERG}) admits a couple of free parameters consistent with
locality.  We tune them to satisfy (\ref{sec6-W-WT}) as much as
possible.  What is left is the anomaly.

At the end of Sec.~II, the current was derived as
\begin{equation}
  J_\mu^a (p) = \int_q \bar{\psi} (-q) T^a \gamma_\mu a_R \psi (q+p)\,.
\end{equation}
The counterterms $\PP$ are quadratic in fields, and we can write them in the form
\begin{align}
  \PP_{\mu_1\cdots\mu_n}^{a_1\cdots a_n} (p_1,\cdots,p_n)
  &= \int_q \bar{\psi} (-q) c_{\mu_1\cdots\mu_n}^{a_1\cdots a_n}
  (p_1,\cdots,p_n; -q, q+p_1+\cdots+p_n) \psi (q+p_1+\cdots +p_n)\nt\\
&\quad  + d_{\mu_1\cdots\mu_n}^{a_1\cdots a_n} (p_1,\cdots,p_n)
  \,\delta \left(\sum_{i=1}^n p_i\right)\,,
\end{align}
where
\begin{align}
  &c_{\mu_1\cdots\mu_n}^{a_1\cdots a_n} (p_1,\cdots,p_n; -q,
    q+p_1+\cdots+p_n)\nt\\
  &= \sum_{\sigma \in \mathcal{S}_n} T^{a_{\sigma (1)}} \cdots T^{a_{\sigma (n)}}
    \gamma_{\mu_{\sigma (1)}} h_F (q+p_{\sigma
    (1)})\gamma_{\mu_{\sigma (2)}} h_F (q+p_{\sigma (1)}+p_{\sigma
    (2)})  \nt\\ 
  &\qquad \cdots \gamma_{\mu_{\sigma (n-1)}} h_F (q+p_{\sigma (1)} + \cdots
    +p_{\sigma (n-1)}) \gamma_{\mu_{\sigma (n)}}\\
  &= \sum_{\sigma \in \mathcal{S}_n} \includegraphics{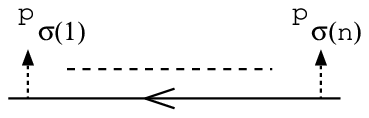}\,.\nt
\end{align}
The sum is taken over all the permutations of $1, \cdots, n$.

Similarly, we can write
\begin{align}
&  d_{\mu_1 \cdots \mu_n}^{a_1 \cdots a_n} (p_1, \cdots, p_n)\nt\\
  &= \sum_{\sigma \in \mathcal{S}_{n-1}}
  \Tr \left( T^{a_1} T^{a_{\sigma (2)}} \cdots T^{a_{\sigma
    (n-1)}} T^{a_{\sigma (n)}} \right)
\cdot d_{\mu_1 \mu_{\sigma (2)} \cdots \mu_{\sigma
      (n)}} (p_1, p_{\sigma (2)} , \cdots,  p_{\sigma(n)})\,,
\end{align}
where the sum is taken over all the permutations of $2, \cdots, n$.  $d$'s
satisfy the ERG equations
\begin{align}
&\left( \sum_{i=1}^n p_i \cdot \partial_{p_i} + n - 4 \right)
                 d_{\alpha_1 \cdots \alpha_n} (p_1, \cdots, p_n)\nt\\
  &= (-) \int_q \Tr f_F (q) \left[ \gamma_{\alpha_1} h_F (q+p_1)
    \gamma_{\alpha_2} \cdots\gamma_{\alpha_{n-1}} h_F (q+p_1+\cdots + p_{n-1})
    \gamma_{\alpha_n} \right.\nt\\
  &\left.\qquad + \gamma_{\alpha_2} h_F (q+p_2) \gamma_{\alpha_3}
    \cdots \gamma_{\alpha_{n}} h_F
    (q+p_2+\cdots + p_n) \gamma_{\alpha_1} + \cdots \right]\,,\label{sec6-naive-ERG}
\end{align}
where
\begin{subequations}
  \begin{align}
    h (p) &\equiv \frac{1-K(p)}{p^2}\,,\\
    f (p) &\equiv (p \cdot \partial_p + 2) h(p) = \frac{\Delta (p)}{p^2}\,,\\
    f_F (p) &\equiv f(p) a_R \fs{p} = a_R \frac{\Delta (p)}{\fs{p}}\,.
  \end{align}
\end{subequations}
For $n \ge 5$, the solutions are given by the finite loop integrals:
\begin{align}
&  d_{\mu_1\cdots\mu_n} (p_1, \cdots, p_n)\nt\\
  &= (-) \int_q \Tr  \left[ \gamma_{\mu_1} h_F (q+p_1)  \gamma_{\mu_2}
    h_F (q+p_1+p_{2}) \cdots h_F (q-p_n) \gamma_{\mu_n} h_F (q)
    \right]\label{sec6-d-large-n}\\  
  &=\raisebox{-1cm}{\includegraphics[height=3cm]{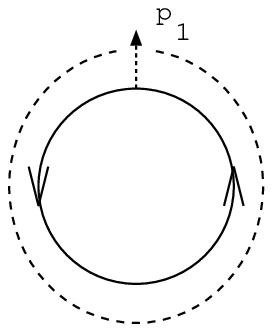}}\,.\nt
\end{align}
For $n=2,3,4$, however, the above loop integrals are UV divergent, and
we must define $d$'s as solutions of (\ref{sec6-naive-ERG}).  We
emphasize that there is no need to introduce an additional UV cutoff
to regularize the loop integrals.  In fact we need to modify
(\ref{sec6-naive-ERG}) first by adding local terms proportional to the
coefficient $b$ so that the solutions become analytic at zero momenta.
ERG then determines $d_2$ up to $t$-independent terms quadratic in
momenta, $d_3$ up to terms linear in momenta, and $d_4$ up to a
constant.  To remove the ambiguities we can resort to the WT identity,
which would be given by
\begin{align}
&  p_{1\alpha} d_{\alpha \alpha_2 \cdots \alpha_n} (p_1, \cdots, p_n)
= d_{\alpha_2 \cdots \alpha_n} (p_1+p_2, p_3, \cdots, p_n) -
    d_{\alpha_2 \cdots \alpha_n} (p_2, \cdots, p_{n-1}, p_n+p_1)\nt\\
  &\quad + \int_q K(q) \Tr \left[
    h_F (q-p_1) \gamma_{\alpha_2} h_F (q+p_2) \cdots h_F (q+p_2+\cdots +
    p_{n-1}) \gamma_{\alpha_n} \right.\nt\\
  &\qquad\quad\left. - h_F (q+p_1) \gamma_{\alpha_2} h_F (q+p_1+p_2)
    \cdots h_F (q+p_1+\cdots + p_{n-1}) \gamma_{\alpha_n} \right]\,,
\end{align}
if there were no anomaly.  This is satisfied by (\ref{sec6-d-large-n})
for $n \ge 5$.  But this is corrected for $n=3, 4$ by the anomaly,
proportional to $\mathcal{A}$.  We can obtain $\mathcal{A}$ by
expanding the WT identity in powers of small momenta.  This is a
straightforward calculation.

In the following we sketch the calculation of
$d_{\alpha_1 \cdots \alpha_n}$ for $n=2,3,4$.  The case $n=2$ is
sufficient to determine the coefficient $b$, but we need the case
$n=3$ to determine $\mathcal{A}$.  We calculate the case $n=4$ for
completeness and check of our formalism.  The essential steps are
expansions of $d$'s in small momenta.  The calculations are all
straightforward, and thanks to the presence of a finite cutoff there
is no hidden subtlety.  Perhaps we could have condensed this section
into a smaller number of pages, but we have decided to give all the
details for the reader unfamiliar with calculations with cutoff functions.
The more experienced reader may skip what seems trivial or redundant.

\subsection{Product of Two $n=2$}

$d_{\alpha\beta}^{a b} (p_1, p_2) = \delta_{a b} d_{\alpha\beta} (p,-p)$ satisfies the ERG
equation
\begin{align}
  & \left( p \cdot \partial_p - 2 \right) d_{\alpha\beta} (p,-p)\nt\\
  &= (-) \int_q \Tr f_F (q) \left( \gamma_\alpha h_F (q+p) \gamma_\beta +
    \gamma_\beta h_F (q-p) \gamma_\alpha \right)
    + b (p^2 \delta_{\alpha\beta} - p_\alpha p_\beta)\,,\label{sec6-two-ERG}
\end{align}
and the Ward identity
\begin{equation}
  p_\alpha d_{\alpha \beta} (p,-p)
 = \int_q K(q) \Tr \left[ h_F (q-p) \gamma_\beta - h_F (q+p)
   \gamma_\beta \right]\,.\label{sec6-two-WT}
\end{equation}

The analyticity of $d_{\alpha\beta} (p,-p)$ at $p=0$ demands that the
rhs (\ref{sec6-two-ERG}) be free of quadratic terms in $p$.  (If there
were, we would obtain a nonlocal $p^2 \ln p$ dependence.)  To expand
the integral on the rhs of (\ref{sec6-two-ERG}) in powers of $p$, we
use
\begin{equation}
  \Tr a_R \fs{a} \fs{b} \fs{c} \fs{d} = 2 \left[ (ab)(cd) + (ad)(bc) -
    (ac)(bd) + \ep_{\alpha\beta\gamma\delta} a_\alpha b_\beta c_\gamma
    d_\delta \right]\,,
\end{equation}
where $\ep_{1234} = 1$, and
\begin{equation}
  h(q+p) = h(q) + (2 (qp) + p^2) h' (q) + \frac{1}{2} (2 (qp))^2 h''
  (q) + \mathrm{O} (p^3)\,,
\end{equation}
where $h' (q) \equiv \frac{d}{dq^2} h(q)$, etc.  We obtain
\begin{align}
  &  (-) \int_q \Tr f_F (q) \left( \gamma_\alpha h_F (q+p) \gamma_\beta +
    \gamma_\beta h_F (q-p) \gamma_\alpha \right)\nt\\
  &\quad\overset{p \to 0}{\longrightarrow}
2 \delta_{\alpha\beta} \int_q f(q) h(q) q^2   - b_2 (p^2
    \delta_{\alpha\beta} - p_\alpha p_\beta )\,,
\end{align}
where
\begin{equation}
  b_2 \equiv - 4 \int_q f(q) \left( q^2 h' (q) + \frac{1}{3} q^4 h''
        (q)\right) = \frac{1}{(4 \pi)^2} \frac{4}{3}\,.\label{sec6-b2}
\end{equation}
The integrand is a total derivative, and the value of the integral is
independent of the choice of the cutoff function $K(p)$.  (See SubSec.~3 of
Appendix \ref{appendix-integrals} for the calculation.)

Hence, with the choice
\begin{equation}
  b =  b_2 = \frac{1}{(4 \pi)^2} \frac{4}{3}\,,\label{sec6-b}
\end{equation}
the general solution of (\ref{sec6-two-ERG}) is given by
\begin{align}
  d_{\alpha\beta} (p, -p)
  &=   - \delta_{\alpha\beta} \int_q f(q) h(q) q^2\nt\\
  &\quad + \int_{-\infty}^0 dt \, e^{- 2 t} \left[
(-)  \int_q \Tr f_F (q) \left( \gamma_\alpha h_F (q+ p e^t)
    \gamma_\beta + \gamma_\beta h_F (q-p e^t) \gamma_\alpha
    \right)\right.\nt\\
  &\left.\qquad + \int_q \Tr f_F (q) \left(\gamma_\alpha h_F (q) \gamma_\beta +
    (\alpha \leftrightarrow \beta) \right)\nt + 4 b \left( p^2
    \delta_{\alpha\beta} - p_\alpha p_\beta  \right) e^{2t} \right]\nt\\
  &\quad + A p^2 \delta_{\alpha\beta} + B \left( p_\alpha p_\beta -
    p^2 \delta_{\alpha\beta} \right)\,,\label{sec6-two-ERG-solution}
\end{align}
where $A, B$ are free parameters.  The subtractions make the integrand
of order $e^{2t}$ as $t \to - \infty$, and the integral is convergent.

We can fix $A$ using the WT identity (\ref{sec6-two-WT}).  First note
\begin{equation}
  p_\alpha d_{\alpha\beta} (p,-p) \overset{p \to 0}{\longrightarrow} -
  p_\alpha d_{\alpha\beta} (0,0) + A\, p^2 p_\beta\,.\label{sec6-two-quadratic}
\end{equation}
To determine $A$ we compute the rhs of (\ref{sec6-two-WT}):
\begin{align}
  & \int_q K(q) \Tr \left( h_F (q-p) \gamma_\beta - h_F (q+p)
    \gamma_\beta \right) = - 2 \int_q K(q) h(q+p) \Tr \left( \fs{q} + \fs{p} \right)
    \gamma_\beta a_R \nt\\
  &\overset{p \to 0}{\longrightarrow}
  p_\beta  (-4) \int_q K(q) \left( h(q)+\frac{1}{2} q^2 h' (q)\right)\nt\\
&\qquad + p^2 p_\beta (-4) \int_q K(q) \lb h' (q) + q^2 h'' (q) + \frac{1}{6}
    q^4 h''' (q) \rb\,.
\end{align}
Consistency with (\ref{sec6-two-quadratic}) demands
\begin{equation}
  \int_q f(q) h(q) q^2 = \int_q K(q) \left( 4 h(q) + 2 q^2 h' (q)
  \right)\,,\label{sec6-integral}
\end{equation}
and
\begin{equation}
A =  - 4 \int_q K(q) \lb h' (q) + q^2 h'' (q) + \frac{1}{6}
q^4 h''' (q) \rb\,.
\end{equation}
The first equation must hold since the WT identity is an operator
equation consistent with ERG.  We verify it explicitly in SubSec.~5 of
Appendix \ref{appendix-integrals}.
There, in SubSec.~4, we also compute
\begin{equation}
A = \frac{1}{(4 \pi)^2} \frac{1}{3} \,.\label{sec6-A}
\end{equation}
$B$ is left arbitrary.

Let us stop here to examine the asymptotic behavior of
$d_{\alpha\beta} (p,-p)$ for large $p$.  In principle we could obtain
the asymptotic behavior using the solution
(\ref{sec6-two-ERG-solution}).  Instead, it is easier to go back to (\ref{sec6-two-ERG})
and (\ref{sec6-two-WT}), which give
\begin{align}
  \left( p \cdot \partial_p - 2 \right) d_{\alpha\beta} (p, -p)
  &\overset{p \to \infty}{\longrightarrow} b \left(p^2 \delta_{\alpha\beta}
    - p_\alpha p_\beta \right)\,,\\
    p_\alpha d_{\alpha\beta} (p,-p) &\overset{p \to
                                      \infty}{\longrightarrow} 0\,.
\end{align}
Hence, we obtain the asymptotic behavior
\begin{equation}
  d_{\alpha\beta} ( p e^t, - p e^t) \overset{t \to
    +\infty}{\longrightarrow}
  b \, t \, e^{2t} \left(p^2
    \delta_{\alpha\beta} - p_\alpha p_\beta\right)\,,
\end{equation}
determined by the constant $b$.  Using this we can construct
the continuum limit as\cite{Sonoda:2017rro}
\begin{equation}
  D_{\alpha\beta} (p, -p) \equiv \lim_{t \to \infty} e^{-2t} \left(
    d_{\alpha\beta} (p e^t, - p e^t) - b t e^{2 t} \left( p^2
      \delta_{\alpha\beta} - p_\alpha p_\beta \right) \right)\,.
\end{equation}
This satisfies
\begin{equation}
  \left( p \cdot \partial_p - 2 \right) D_{\alpha\beta} (p, -p) = b
  \left( p^2 \delta_{\alpha\beta} - p_\alpha p_\beta \right)\,.
\end{equation}
Since $D_{\alpha\beta}$ depends on the constant $B$, we can rewrite
this as
\begin{equation}
  \left( p \cdot \partial_p - 2 + b \partial_B \right) D_{\alpha\beta} (p, -p) = 0\,.
\end{equation}
$D_{\alpha\beta} (p,-p)$ is also transverse:
\begin{equation}
  p_\alpha D_{\alpha\beta} (p,-p) = 0\,.
\end{equation}

The two-point function of the current is now obtained as
\begin{equation}
  \vvev{J_\alpha^a (p) J_\beta^b (q)}_B
  = \delta^{ab} \delta (p+q) \,D_{\alpha\beta} (p, -p)\,,
\end{equation}
which is transverse, and satisfies the scaling relation
\begin{equation}
  \left( p \cdot \partial_p + q \cdot \partial_q + 2 + b
\partial_B \right) \vvev{J_\alpha^a (p) J_\beta^b
    (q)}_B = 0\,.
\end{equation}

\subsection{Product of Three $n=3$}

$d_{\alpha\beta\gamma} (p_1, p_2, p_3)$ satisfies the ERG equation
\begin{align}
  &\left( \sum_{i=1}^3 p_i \cdot \partial_{p_i} - 1 \right)
  d_{\alpha\beta\gamma} (p_1,p_2,p_3)\nt\\
  &= (-) \int_q \Tr f_F (q) \left[ \gamma_\alpha h_F (q+p_1)
    \gamma_\beta h_F (q+p_1+p_2) \gamma_\gamma \right.\nt\\
  &\quad\left. + \gamma_\beta h_F (q+p_2) \gamma_\gamma h_F
    (q+p_2+p_3) \gamma_\alpha + \gamma_\gamma h_F (q+p_3)
    \gamma_\alpha h_F (q+p_3+p_1) \gamma_\beta \right] \nt\\
  &\quad + b \left[ \delta_{\alpha\beta} (p_1-p_2)_\gamma +
    \delta_{\beta\gamma} (p_2-p_3)_\alpha + \delta_{\gamma\alpha}
    (p_3-p_1)_\beta \right]\,,\label{sec6-three-ERG}
\end{align}
where $b$ is given by (\ref{sec6-b}), and the WT identity
\begin{align}
  &p_{1\alpha} d_{\alpha\beta\gamma} (p_1, p_2, p_3)
  = d_{\beta\gamma} (p_1+p_2, p_3) - d_{\beta\gamma} (p_2, p_3+p_4)\nt\\
  &\quad + \int_q K(q) \Tr \left[ h_F (q-p_1) \gamma_\beta h_F (q+p_2)
    \gamma_\gamma - h_F (q+p_1) \gamma_\beta h_F (q+p_1+p_2)
    \gamma_\gamma \right]\nt\\
  &\quad - \frac{1}{2} \mathcal{A} \ep_{\alpha\beta\gamma\delta}
    p_{1\alpha} \left( p_2 - p_3 \right)_\delta\,,\label{sec6-three-WT}
\end{align}
where $\mathcal{A}$ is to be determined.

Analyticity of $d_{\alpha\beta\gamma}$ at $p_i=0$ requires the absence
of terms linear in $p_i$ from the rhs of (\ref{sec6-three-ERG}).
($p_i$ would imply nonlocal $p_i \ln p_j$.)  Let
us check it.  Expanding
the integral in momenta, we obtain the linear terms as
\begin{equation}
  (-) 2 \int_q f(q) h(q)^2 q^2 \left[ \delta_{\alpha\beta} (p_1 -
    p_2)_\gamma + \delta_{\beta\gamma} (p_2-p_3)_\alpha +
    \delta_{\gamma\alpha} (p_3-p_1)_\beta \right]\,.
\end{equation}
The integrand is a total derivative, and we obtain
\begin{equation}
 - 2 \int_q f(q) h(q)^2 q^2 = - \frac{1}{(4 \pi)^2} \frac{4}{3} = - b\,.
\end{equation}
(See SubSec.~1 of Appendix \ref{appendix-integrals}.)
Hence, (\ref{sec6-three-ERG}) is consistent with locality.

The general solution is given by
\begin{align}
  &  d_{\alpha\beta\gamma} (p_1, p_2, p_3)\nt\\
  &= \int_{-\infty}^0 dt\, e^{-t} \left[  (-) \int_q \Tr f_F (q)
    \lb\gamma_\alpha h_F (q+p_1 e^t)
    \gamma_\beta h_F (q+(p_1+p_2)e^t) \gamma_\gamma \right.\right.\nt\\
  &\left.\qquad\quad + \gamma_\beta h_F (q+p_2 e^t) \gamma_\gamma h_F
    (q+(p_2+p_3)e^t) \gamma_\alpha + \gamma_\gamma h_F (q+p_3 e^t)
    \gamma_\alpha h_F (q+(p_3+p_1)e^t) \gamma_\beta \rb\nt\\
  &\quad\qquad - b \lb \delta_{\alpha\beta}
    (p_1-p_2)_\gamma + \delta_{\beta\gamma} (p_2-p_3)_\alpha +
    \delta_{\gamma\alpha} (p_3-p_1)_\beta \rb e^t \bigg] \nt\\
  &\quad  +  c_{\alpha\beta\gamma\delta}
  p_{1 \delta} + c_{\beta\gamma\alpha\delta} p_{2\delta} +
  c_{\gamma\alpha\beta\delta} p_{3\delta}\,,\label{sec6-three-ERG-solution}
\end{align}
where $c_{\alpha\beta\gamma\delta}$ are arbitrary constants, not
determined by (\ref{sec6-three-ERG}).  We note that the integrand
behaves as $e^t$ as $t \to - \infty$, and the integral is convergent.
The particular form of the linear terms is required by the cyclic
symmetry:
\begin{equation}
  d_{\alpha\beta\gamma} (p_1, p_2, p_3) = d_{\beta\gamma\alpha} (p_2,
  p_3, p_1) = d_{\gamma\alpha\beta} (p_3, p_1, p_2)\,.
\end{equation}
The most general form of $c_{\alpha\beta\gamma\delta}$ is given by
\begin{equation}
c_{\alpha\beta\gamma\delta} = s\, \delta_{\alpha\beta}
\delta_{\gamma\delta} + t\, \delta_{\alpha\gamma} \delta_{\beta\delta}
+ u\, \delta_{\alpha\delta} \delta_{\beta\gamma} \,,
\end{equation}
where $s, t, u$ are constants.\footnote{Cyclic symmetry allows a term
  proportional to $\ep_{\alpha\beta\gamma\delta}$, but it does not
  contribute to $d_{\alpha\beta\gamma}$.  Similarly,
  $c_{\alpha\beta\gamma\delta}$ does not change if we change $s, t, u$
  by the same amount.  So, we could set $u$ to zero.}

We now wish to show that we can choose $s, t, u$, and $\mathcal{A}$ so
that (\ref{sec6-three-WT}) is valid.  Since (\ref{sec6-three-WT}) is
consistent with (\ref{sec6-three-ERG}), we only need to check the
terms quadratic in momenta.  Using
\begin{equation}
  d_{\alpha\beta} (p,-p) \overset{p \to 0}{\longrightarrow}
  - \delta_{\alpha\beta} \int_q f(q) h(q) q^2 + A p^2
  \delta_{\alpha\beta} +  B \left(p_\alpha p_\beta - p^2
    \delta_{\alpha\beta}\right)\,,
\end{equation}
we obtain
\begin{equation}
  d_{\beta\gamma} (p_1+p_2, p_3) - d_{\beta\gamma} (p_2, p_3+p_1)
  \overset{p_i \to 0}{\longrightarrow} A (p_3^2-p_2^2)
  \delta_{\beta\gamma} + B \lb
  p_{3\beta} p_{3\gamma} - p_3^2 \delta_{\beta\gamma} - p_{2\beta}
  p_{2\gamma} + p_2^2 \delta_{\beta\gamma} \rb\,,
\end{equation}
where $A$ is given by (\ref{sec6-A}).

We next consider the small momentum behavior of the integral on the
rhs of (\ref{sec6-three-WT}):
\begin{align}
  & \int_q K(q) \Tr \left[ h_F (q-p_1) \gamma_\beta h_F (q+p_2) - h_F
    (q+p_1) \gamma_\beta h_F (q-p_3) \right] \gamma_\gamma \nt\\
  &= \int_q K(q) \Tr \left[ h_F (q+p_1) \gamma_\beta \left( h_F
    (q-p_2) - h_F (q-p_3) \right) \gamma_\gamma\right]\nt\\
  &= \int_q K(q) \left[ h(q+p_1) h(q-p_2) \Tr a_R
    (\fs{q}+\fs{p}_1)\gamma_\beta (\fs{q} - \fs{p}_2) \gamma_\gamma
    \right.\nt\\
  &\left.\qquad\qquad-
    h(q+p_1) h(q-p_3) \Tr a_R (\fs{q} + \fs{p}_1)\gamma_\beta (\fs{q}
    - \fs{p}_3) \gamma_\gamma \right]\nt\\
  &\overset{p_i \to 0}{\longrightarrow}
    \left( - 2 \ep_{\alpha\beta\delta\gamma}
    p_{1 \alpha} (p_2-p_3)_\delta  - 2 (p_3^2 - p_2^2)
    \delta_{\beta\gamma} \right) \int_q K(q) \left( h(q)^2 + h (q) q^2
    h'(q) \right)\nt\\ 
  &\quad + \left( p_{3\beta} p_{3\gamma} - p_3^2 \delta_{\beta\gamma}
    - p_{2\beta} p_{2\gamma} + p_2^2 \delta_{\beta\gamma} \right)\nt\\
  &\qquad\qquad\qquad \times 
    \int_q K(q) \left( - 4 h(q)^2 - 6 h(q) q^2 h'(q) - \frac{4}{3}
    (q^2 h'(q))^2 - \frac{2}{3} h (q) q^4 h'' (q) \right)\,,
\end{align}
where the first integral, whose integrand is a total derivative, can be calculated as
\begin{equation}
  \int_q K(q) \left( h(q)^2 + h(q) q^2 h'(q) \right) = \frac{1}{(4
    \pi)^2} \frac{1}{6}\,.
\end{equation}
(See SubSec.~2 of Appendix \ref{appendix-integrals}.)  Hence, we
obtain the rhs of (\ref{sec6-three-WT}) as
\begin{align}
  \mathrm{rhs}
  &\overset{p_i \to 0}{\longrightarrow}
    \left( p_{3\beta} p_{3\gamma} - p_3^2 \delta_{\beta\gamma} -
    p_{2\beta} p_{2\gamma} + p_2^2 \delta_{\beta\gamma} \right) \nt\\
  &\qquad\qquad \times \left[
    B - \int_q K \left( 4 h^2 + 6 h q^2 h' + \frac{4}{3} (q^2 h')^2 +
    \frac{2}{3} h q^4 h'' \right) \right]\nt\\
  &\qquad + \left(\frac{1}{(4 \pi)^2} \frac{1}{3} - \frac{1}{2} \mathcal{A}\right)
    \ep_{\alpha\beta\gamma\delta} p_{1\alpha} (p_2-p_3)_\delta\,.\label{sec6-three-WT-rhs}
\end{align}

We next compute the small momentum behavior of the lhs of
(\ref{sec6-three-WT}).  From
\begin{align}
&  d_{\alpha\beta\gamma} (p_1, p_2, p_3)\nt\\
& \overset{p_i \to 0}{\longrightarrow} \delta_{\alpha\beta} (s\, p_1 +
  t\, p_2 + u p_3)_\gamma + \delta_{\beta\gamma} (u\, p_1 + s\, p_2 +
  t\, p_3)_\alpha + \delta_{\gamma\alpha} (t\, p_1 + u\, p_2 + s\,
  p_3)_\beta\,,
\end{align}
we obtain
\begin{align}
  p_{1 \alpha} d_{\alpha\beta\gamma} (p_1, p_2, p_3)
  &\overset{p_i \to 0}{\longrightarrow} p_{1\alpha} \left(c_{\alpha\beta\gamma\delta}
    p_{1\delta} + c_{\beta\gamma\alpha\delta} p_{2\delta} +
    c_{\gamma\alpha\beta\delta} p_{3 \delta} \right)\nt\\
&= \left(p_{2\beta} p_{2\gamma} - \delta_{\beta\gamma} p_2^2 \right) (s-u)
+  \left(p_{3\beta} p_{3\gamma}  - p_3^2\delta_{\beta\gamma}
                                                         \right)(t-u)\nt  \\
  &\quad + (s+t-2u) \left( p_{2\beta} p_{3\gamma} -
    \delta_{\beta\gamma} (p_2 p_3) \right)\,.
\end{align}
Matching this with (\ref{sec6-three-WT-rhs}), we obtain
\begin{subequations}
\begin{align}
  u & = \frac{1}{2} (s+t)\,,\\
  \frac{1}{2} (s-t)
  &= - B + \int_q K(q) \left( 4 h^2 + 6 h q^2 h' + \frac{4}{3} (q^2
    h')^2 + \frac{2}{3} h q^4 h''
    \right)\,,\label{sec6-three-sminust}
\end{align}
\end{subequations}
which determine the low momentum behavior
\begin{equation}
  d_{\alpha\beta\gamma} (p_1,p_2,p_3) \overset{p_i \to
    0}{\longrightarrow}
  \frac{1}{2} (s-t) \left( \delta_{\alpha\beta} (p_1-p_2)_\gamma +
    \delta_{\beta\gamma} (p_2-p_3)_\alpha + \delta_{\gamma\alpha}
    (p_3-p_1)_\beta \right)\,.\label{sec6-three-smallp}
\end{equation}
We also obtain the coefficient of the anomaly as\footnote{It was first
  pointed out in \cite{Wilson:1969zs} that the chiral anomaly comes from the
  short-distance singularity of three currents.  The calculation
  following this suggestion was completed in coordinate space in
  \cite{Sonoda:1996fi}.}
\begin{equation}
  \mathcal{A}
  = \frac{1}{(4 \pi)^2} \frac{2}{3}\,.\label{sec6-calA}
\end{equation}

Let us stop here to examine the asymptotic behavior of
$d_{\alpha\beta\gamma} (p_1, p_2, p_3)$ for large momenta.  Instead of
taking the asymptotic limit of (\ref{sec6-three-ERG-solution}), we go
back to (\ref{sec6-three-ERG}) and (\ref{sec6-three-WT}).
For large momenta, (\ref{sec6-three-ERG}) gives
\begin{align}
  \left(\sum_{i=1}^3 p_i \cdot \partial_{p_i} - 1 \right)
  d_{\alpha\beta\gamma} (p_1, p_2, p_3) \overset{p_i \to
  \infty}{\longrightarrow}
  b \left[ \delta_{\alpha\beta} (p_1 - p_2)_\gamma +
  \delta_{\beta\gamma} (p_2-p_3)_\alpha + \delta_{\gamma\alpha}
  (p_3-p_1)_\beta \right]\,,\label{sec6-three-ERG-asymptotic}
\end{align}
and (\ref{sec6-three-WT}) gives
\begin{equation}
  p_{1\alpha} d_{\alpha\beta\gamma} (p_1, p_2, p_3)
  \overset{p_i \to \infty}{\longrightarrow} d_{\beta\gamma} (-p_3,
  p_3) - d_{\beta\gamma} (p_2, - p_2) - \frac{1}{2} \mathcal{A}
  \ep_{\alpha\beta\gamma\delta} p_{1\alpha}
  (p_2-p_3)_\delta\,.\label{sec6-three-WT-asymptotic}
\end{equation}
(\ref{sec6-three-ERG-asymptotic}) gives the dominant asymptotic
behavior 
\begin{equation}
  d_{\alpha\beta\gamma} (p_1 e^t, p_2 e^t, p_3 e^t)
  \overset{t \to \infty}{\longrightarrow} b\, t\,  e^t \left[
    \delta_{\alpha\beta} (p_1 - p_2)_\gamma + 
  \delta_{\beta\gamma} (p_2-p_3)_\alpha + \delta_{\gamma\alpha}
  (p_3-p_1)_\beta \right]\,,
\end{equation}
which is proportional to the coefficient $b$.  Hence, we can construct
the continuum limit as
\begin{align}
  D_{\alpha\beta\gamma} (p_1, p_2, p_3)
  &\equiv \lim_{t \to +\infty} e^{-t} \Big[ d_{\alpha\beta\gamma} (p_1
    e^t, p_2 e^t, p_3 e^t) \nt\\
  &\qquad - b t e^t \lb \delta_{\alpha\beta} (p_1 - p_2)_\gamma
    + \delta_{\beta\gamma} (p_2-p_3)_\alpha + \delta_{\gamma\alpha}
    (p_3 -p_1)_\beta \rb \Big]\,.
\end{align}
This satisfies the scaling relation
\begin{align}
&  \left( \sum_{i=1}^3 p_i \cdot \partial_{p_i} - 1 \right)
                D_{\alpha\beta\gamma} (p_1, p_2, p_3)\nt\\
  &= b \lb \delta_{\alpha\beta} (p_1 - p_2)_\gamma
    + \delta_{\beta\gamma} (p_2-p_3)_\alpha + \delta_{\gamma\alpha}
    (p_3 -p_1)_\beta \rb \nt\\
  &= - b \partial_B  D_{\alpha\beta\gamma} (p_1, p_2, p_3)\,,
\end{align}
and the WT identity:
\begin{equation}
  p_{1\alpha} D_{\alpha\beta\gamma} (p_1, p_2, p_3) = D_{\beta\gamma}
  (-p_3, p_3) - D_{\beta\gamma} (p_2, -p_2) - \frac{1}{2} \mathcal{A}
  \ep_{\alpha\beta\gamma\delta} p_{1\alpha} \left( p_2 - p_3
  \right)_\delta\,.
\end{equation}

The continuum limit of the connected three-point function defined by
\begin{align}
  &  \vvev{J_\alpha^a (p_1) J_\beta^b (p_2) J_\gamma^c (p_3)}^{conn}_B\nt\\
  &\equiv \delta (p_1+p_2+p_3) \left[ \Tr T^a T^b T^c \,
    D_{\alpha\beta\gamma} (p_1, p_2, p_3) + \Tr T^a T^c T^b\,
    D_{\alpha\gamma\beta} (p_1, p_3, p_2) \right]
\end{align}
satisfies the scaling relation
\begin{equation}
  \left( \sum_{i=1}^3 p_i \cdot \partial_{p_i} + 3 + b \partial_B
  \right)  \vvev{J_\alpha^a (p_1) J_\beta^b (p_2) J_\gamma^c (p_3)}^{conn}_B
  = 0\,,
\end{equation}
and the WT identity
\begin{align}
&  p_{1\alpha} \vvev{J_{\alpha}^{a} (p_1) J_\beta^{b} (p_2)
                J_\gamma^{c} (p_3)}_B^{conn} \nt\\
  &= i f^{abd} \vvev{J_\beta^d (p_1+p_2) J_\gamma^{c} (p_3)}_B^{conn}
    + i f^{acd} \vvev{J_\beta^{b} (p_2) J_\gamma^d
    (p_1+p_3)}^{conn}_B\nt\\  &\quad - \frac{1}{2} \mathcal{A}\, \Tr T^{a}
                           \lb T^{b},  T^{c}\rb \, \ep_{\alpha\beta\gamma\delta}
    p_{1\alpha} (p_2-p_3)_\delta \, \delta (p_1+p_2+p_3)\,.
\end{align}

\subsection{Product of Four $n=4$}

$d_{\alpha\beta\gamma\delta} (p_1, p_2, p_3, p_4)$ must
satisfy the ERG equation
\begin{align}
&  \sum_{i=1}^4 p_i \cdot \partial_{p_i} \, d_{\alpha\beta\gamma\delta}
  (p_1, p_2, p_3, p_4)\nt\\
&= (-) \int_q \Tr f_F (q) \left[ \gamma_\alpha h_F (q+p_1)
                           \gamma_\beta h_F (q+p_1+p_2) \gamma_\gamma
                           h_F (q+p_1+p_2+p_3) \gamma_\delta \right.\nt\\
  &\quad + \gamma_\beta h_F (q+p_2) \gamma_\gamma h_F (q+p_2+p_3)
    \gamma_\delta h_F (q+p_2+p_3+p_4) \gamma_\alpha\nt\\
  &\quad + \gamma_\gamma h_F (q+p_3) \gamma_\delta h_F (q+p_3+p_4)
    \gamma_\alpha h_F (q+p_3+p_4+p_1) \gamma_\beta \nt\\
  &\left.\quad + \gamma_\delta h_F (q+p_4) \gamma_\alpha h_F
    (q+p_4+p_1) \gamma_\beta h_F (q+p_4+p_1+p_2) \gamma_\gamma
    \right]\nt\\
  &\quad + b \left( \delta_{\alpha\beta} \delta_{\gamma\delta} +
    \delta_{\beta\gamma} \delta_{\alpha\delta} - 2
    \delta_{\alpha\gamma} \delta_{\beta\delta}\right)\,,
    \label{sec6-four-ERG}
\end{align}
where $b$ is given by (\ref{sec6-b}), and the WT identity
\begin{align}
  p_{1\alpha} d_{\alpha\beta\gamma\delta} (p_1, p_2, p_3, p_4)
&= d_{\beta\gamma\delta} (p_1+p_2, p_3, p_4) - d_{\beta\gamma\delta}
    (p_2, p_3, p_4+p_1)\nt\\
  &\quad + \int_q K(q) \Tr \left[ h_F (q-p_1) \gamma_\beta h_F (q+p_2)
    \gamma_\gamma h_F (q+p_2+p_3) \right.\nt\\
  &\left.\qquad\quad - h_F (q+p_1) \gamma_\beta h_F
    (q+p_1+p_2) \gamma_\gamma h_F (q+p_1+p_2+p_3) \right]
    \gamma_\delta\nt\\
  &\quad - \frac{1}{2} \mathcal{A} p_{1 \alpha}
    \ep_{\alpha\beta\gamma\delta}\,,\label{sec6-four-WT} 
\end{align}
where $\mathcal{A}$ is given by (\ref{sec6-calA}).

We would like to check two things.  As for (\ref{sec6-four-ERG}), we
would like to check the vanishing of the rhs at zero momenta.  (A
constant would imply nonlocal $\ln p$.)  As for
(\ref{sec6-four-WT}), we would like to check its validity at the first
order in momenta.

The rhs of (\ref{sec6-four-ERG}) gives
\begin{equation}
  (\mathrm{rhs}) \overset{p_i \to 0}{\longrightarrow} \left[ (-) \frac{1}{6}
  \int_q f(q) h(q)^3 q^4 \times 16 + b \right] \left( \delta_{\alpha\beta}
    \delta_{\gamma\delta} + \delta_{\beta\gamma} \delta_{\alpha\delta}
    - 2 \delta_{\alpha\gamma} \delta_{\beta\delta} \right)\,.
\end{equation}
The integrand is a total derivative, and we obtain
\begin{equation}
  \frac{8}{3} \int_q f(q) h(q)^3 q^4 = \frac{1}{(4 \pi)^2} \frac{4}{3} = b\,.
\end{equation}
(See SubSec.~1 of Appendix \ref{appendix-integrals}.)  Hence, the rhs
vanishes at zero momenta as desired.

We now wish to check (\ref{sec6-four-WT}) to first order in momenta.
(\ref{sec6-four-ERG}) determines only the momentum dependence of
$d_{\alpha\beta\gamma\delta}$, but its value at $p_i = 0$ is left
undetermined.  The most general form, consistent with cyclic symmetry,
is
\begin{equation}
  d_{\alpha\beta\gamma\delta} (0,0,0,0) = s_4 \left(
    \delta_{\alpha\beta} \delta_{\gamma\delta} + \delta_{\beta\gamma}
    \delta_{\delta\alpha} \right) + t_4 \, \delta_{\alpha\gamma}
  \delta_{\beta\delta}\,,
\end{equation}
where $s_4, t_4$ are constants so that
\begin{equation}
  p_{1\alpha} d_{\alpha\beta\gamma\delta} (0,0,0,0) = s_4 \left(
    p_{1\beta} \delta_{\gamma\delta} + p_{1\delta}
    \delta_{\beta\gamma} \right) + t_4 p_{1\gamma}
  \delta_{\beta\delta}\,.\label{sec6-four-WT-lhs}
\end{equation}
To compare this with the rhs, we first compute
\begin{equation}
  d_{\beta\gamma\delta} (p_1+p_2, p_3, p_4) - d_{\beta\gamma\delta}
  (p_2, p_3, p_4+p_1)
  \overset{p_i \to 0}{\longrightarrow}
  \frac{1}{2} (s-t) \left(  p_{1\delta} \delta_{\beta\gamma} +
    p_{1\beta} \delta_{\gamma\delta} - 2 p_{1\gamma}
    \delta_{\beta\delta} \right)\,,
\end{equation}
where we have used (\ref{sec6-three-smallp}), and $\frac{1}{2} (s-t)$
is given by (\ref{sec6-three-sminust}).  We next compute
\begin{align}
&  \int_q K(q) \Tr \left[ h_F (q-p_1) \gamma_\beta h_F (q+p_2)
                 \gamma_\gamma h_F (q+p_2+p_3) \right.\nt\\
  &\qquad \left.- h_F (q+p_1)
                 \gamma_\beta h_F (q+p_1+p_2) \gamma_\gamma h_F
                 (q+p_1+p_2+p_3) \right] \gamma_\delta\nt\\
  &= \int_q K(q) \left[ h (q-p_1) h(q+p_2) h(q+p_2+p_3) \Tr
    (\fs{q}-\fs{p}_1) \gamma_\beta (\fs{q} + \fs{p}_2) \gamma_\gamma
    (\fs{q} + \fs{p}_2 + \fs{p}_3) \right.\nt\\
  &\qquad - h(q+p_1) h(q+p_1+p_2) h(q+p_1+p_2+p_3) \nt\\
  &\left.\qquad\qquad \times \Tr (\fs{q} +
    \fs{p}_1) \gamma_\beta (\fs{q} + \fs{p}_1 + \fs{p}_2)
    \gamma_\gamma (\fs{q} + \fs{p}_1 + \fs{p}_2 + \fs{p}_3 ) \right]
    \gamma_\delta a_R\nt\\
  &\overset{p_i \to 0}{\longrightarrow}
p_{1\alpha} \ep_{\alpha\beta\gamma\delta} 4 \int_q K(q) h(q)^2 q^2 \left(
    h(q) + q^2 h' (q) \right)\nt\\
  &\qquad + \left(p_{1\beta} \delta_{\gamma\delta} + p_{1\delta}
    \delta_{\beta\gamma}\right) \int_q K(q) q^2 h(q)^2 \left(2 h(q) +
    \frac{4}{3} q^2 h'(q) \right)\nt\\
  &\qquad + p_{1\gamma} \delta_{\beta\delta} \frac{4}{3} \int_q K(q)
    h(q)^2 q^4 h' (q)\,.
\end{align}
Hence, the rhs of (\ref{sec6-four-WT}) is
\begin{align}
  \mathrm{rhs} &\overset{p_i \to 0}{\longrightarrow}
\frac{1}{2}  (s-t) \left( p_{1\beta} \delta_{\gamma\delta} +
  p_{1\delta} \delta_{\beta\gamma} - 2  p_{1\gamma}
  \delta_{\beta\delta} \right)\nt\\
  &\quad + (p_{1\beta} \delta_{\gamma\delta} + p_{1\delta}
    \delta_{\beta\gamma}) \int_q K (q) h(q)^2 q^2 \left(2 h (q) +
    \frac{4}{3}  q^2 h' (q) \right)\nt\\
  &\quad + p_{1\gamma} \delta_{\beta\delta} \frac{4}{3} \int_q K (q)
    h (q)^2 q^4 h' (q)\nt\\
  &\quad + p_{1\alpha} \ep_{\alpha\beta\gamma\delta} \left( 4 \int_q K(q)
    h(q)^2 q^2 \left( h(q) + q^2 h' (q) \right)  - \frac{1}{2}
    \mathcal{A} \right)\,.\label{sec6-four-WT-rhs}
\end{align}
The last term vanishes because
\begin{equation}
  \int_q K(q) h(q)^2 q^2 \left( h(q) + q^2 h' (q) \right) =
  \frac{1}{(4\pi)^2} \frac{1}{12}\,.
\end{equation}
(See SubSec.~2 of Appendix \ref{appendix-integrals}.)  We can make
(\ref{sec6-four-WT-rhs}) match with (\ref{sec6-four-WT-lhs}) by
choosing
\begin{subequations}
\begin{align}
  s_4 &= \frac{1}{2} (s-t) + \int_q K (q) h(q)^2 q^2 \left(2 h (q) +
        \frac{4}{3}  q^2 h' (q) \right)\\
  &= - B + \int_q K \left( 4 h^2 + 6 h q^2 h' + \frac{4}{3} (q^2 h')^2
  + \frac{2}{3} h q^4 h'' + 2 h^3 q^2 + \frac{4}{3} h^2 q^4 h'
    \right)\nt\,, \\
  t_4 &= - (s-t) + \frac{4}{3}  \int_q K (q) h (q)^2 q^4 h' (q)\nt\\
  &= - 2 s_4 + 4 \int_q K(q) h(q)^2 q^2 \left( h(q) + q^2 h' (q)\right)
= - 2 s_4 + \frac{1}{(4 \pi)^2} \frac{1}{3}\,.
\end{align}
\end{subequations}
We have thus checked the validity of (\ref{sec6-four-WT}).

Finally we examine the asymptotic behavior of $d_{\alpha\beta\gamma\delta}
(p_1, p_2, p_3, p_4)$ for large momenta.  (\ref{sec6-four-ERG}) and
(\ref{sec6-four-WT}) give
\begin{align}
  \sum_{i=1}^4 p_i \cdot \partial_{p_i} d_{\alpha\beta\gamma\delta}
  (p_1, p_2, p_3, p_4) &\overset{p_i \to \infty}{\longrightarrow} b
                         \left( \delta_{\alpha\beta}
                         \delta_{\gamma\delta} + \delta_{\beta\gamma}
                         \delta_{\alpha\delta} - 2
                         \delta_{\alpha\gamma} \delta_{\beta\delta}
                         \right)\,,\\
  p_{1 \alpha} d_{\alpha\beta\gamma\delta} (p_1, p_2, p_3, p_4)
  &\overset{p_i \to \infty}{\longrightarrow} d_{\beta\gamma\delta}
    (- p_3-p_4, p_3, p_4) - d_{\beta\gamma\delta} (p_2, p_3, - p_2 -
    p_3) \nt\\
  &\qquad\quad - \frac{1}{2} \mathcal{A} p_{1\alpha}
    \ep_{\alpha\beta\gamma\delta}\,.
\end{align}
The first equation gives the asymptotic behavior
\begin{equation}
  d_{\alpha\beta\gamma\delta} (p_1 e^t, p_2 e^t, p_3 e^t, p_4 e^t)
  \overset{t \to \infty}{\longrightarrow} b \, t \left( \delta_{\alpha\beta}
                         \delta_{\gamma\delta} + \delta_{\beta\gamma}
                         \delta_{\alpha\delta} - 2
                         \delta_{\alpha\gamma} \delta_{\beta\delta} \right)\,.
\end{equation}
Hence, a continuum limit is obtained as
\begin{align}
  &  D_{\alpha\beta\gamma\delta} (p_1, p_2, p_3, p_4)\nt\\
  &\equiv \lim_{t \to +\infty} \left[ d_{\alpha\beta\gamma\delta} (p_1
    e^t, p_2 e^t, p_3 e^t, p_4 e^t) - b t \left(  \delta_{\alpha\beta}
                         \delta_{\gamma\delta} + \delta_{\beta\gamma}
                         \delta_{\alpha\delta} - 2
                         \delta_{\alpha\gamma} \delta_{\beta\delta}
    \right) \right]\,,
\end{align}
which satisfies the scaling relation
\begin{align}
  \sum_{i=1}^4 p_i \cdot \partial_{p_i} \, D_{\alpha\beta\gamma\delta}
  (p_1, p_2, p_3, p_4)
  &= b \left(  \delta_{\alpha\beta}
                         \delta_{\gamma\delta} + \delta_{\beta\gamma}
                         \delta_{\alpha\delta} - 2
                         \delta_{\alpha\gamma}
    \delta_{\beta\delta}\right)\nt\\
  &= - b \partial_B D_{\alpha\beta\gamma\delta} (p_1, p_2, p_3,
    p_4)\,,
\end{align}
and the WT identity
\begin{align}
  p_{1 \alpha} D_{\alpha\beta\gamma\delta} (p_1, p_2, p_3, p_4)
  &= D_{\beta\gamma\delta} (-p_3-p_4, p_3, p_4) - D_{\beta\gamma\delta}
    (p_2, p_3, - p_2 - p_3) \nt\\
  &\quad - \frac{1}{2} \mathcal{A} p_{1\alpha}
  \ep_{\alpha\beta\gamma\delta}\,.
\end{align}
Hence, the connected four-point function defined by
\begin{align}
  \vvev{J_\alpha^a (p_1) J_\beta^b (p_2) J_\gamma^c (p_3) J_\delta^d
    (p_4)}^{conn}_B
 & \equiv \delta (p_1+p_2+p_3+p_4)\, \left[ \Tr T^a T^b T^c T^d \,
   D_{\alpha\beta\gamma\delta} (p_1, p_2, p_3, p_4)\right.\nt\\
  &\left.\quad + \Tr T^a T^b T^d T^c\, D_{\alpha\beta\delta\gamma} (p_1,
    p_2, p_4, p_3) + \cdots \right]
\end{align}
satisfies the scaling relation
\begin{equation}
\left(  \sum_{i=1}^4 p_i \cdot \partial_{p_i} + 4 + b \partial_B
\right) \vvev{J_\alpha^a (p_1) J_\beta^b (p_2) J_\gamma^c (p_3) J_\delta^d
  (p_4)}^{conn}_B = 0
\end{equation}
and the WT identity
\begin{align}
&  p_{1\alpha}  \vvev{J_\alpha^a (p_1) J_\beta^b (p_2) J_\gamma^c (p_3) J_\delta^d
    (p_4)}_B^{conn}
                = i f^{a b e} \vvev{J_\beta^e (p_1+p_2) J_\gamma^c
                (p_3) J_\delta^d (p_4)}_B^{conn} + \cdots\nt\\
&\quad      - \frac{1}{2} \mathcal{A} \, p_{1\alpha}
                \ep_{\alpha\beta\gamma\delta} \,\delta
                                                            (p_1+p_2+p_3+p_4) 
 \times \Tr T^a \left( T^b [T^c , T^d] + T^c [T^d, T^b] + T^d [T^b, T^c]
    \right)\,.
\end{align}

\subsection{Recapitulation}

Let us recapitulate the results of this section by writing down
equations for $e^{W[A]}$, a composite operator of scale dimension $0$.
The ERG differential equation is given by
\begin{align}
  & \left( \int_p \left( - p \cdot \partial_p - D + 1\right) A_\mu^a
    (p) \cdot \frac{\delta}{\delta A_\mu^a (p)} - \mathcal{D}\right)
    e^{W[A]}\nt\\
  &=  \frac{1}{(4\pi)^2} \frac{4}{3}\cdot \frac{1}{4}
    \int d^4 x\, \Tr \left( \partial_\alpha A_\beta - \partial_\beta
    A_\alpha - i [A_\alpha, A_\beta] \right) \left( \partial_\alpha A_\beta - \partial_\beta
    A_\alpha - i [A_\alpha, A_\beta] \right) \,e^{W[A]}\,.
    \end{align}
The WT identity is given by
\begin{align}
\delta_\ep e^{W[A]} &\equiv \int_p \left(  - p_\mu \ep^a (p) + i f^{abc}
  \int_q A_\mu^b (p+q) \ep^c (-q) \right) \frac{\delta}{\delta A_\mu^a
    (p)} \, e^{W[A]}\nt\\
  &= \left[ \int_p \ep^a (-p) \E^a (p) \star + \frac{1}{(4 \pi)^2} \frac{2}{3} \int d^4 x\,
    \ep_{\alpha\beta\gamma\delta} \Tr \partial_\alpha \ep \left(
    A_\beta \partial_\gamma A_\delta - i \frac{1}{2} A_\beta A_\gamma
    A_\delta \right) \right]\, e^{W[A]}\,.
\end{align}

$W[A]$ is determined uniquely by the above two equations up to a
constant multiple of the gauge invariant
\[
  \frac{1}{4} \int d^4 x\, \Tr \left( \partial_\alpha A_\beta - \partial_\beta
    A_\alpha - i [A_\alpha, A_\beta] \right) \left( \partial_\alpha A_\beta - \partial_\beta
    A_\alpha - i [A_\alpha, A_\beta] \right) \,.
\]
If we define
\begin{equation}
  W_g [A] = - \frac{1}{4 g} \int d^4 x\, \Tr \left( \partial_\alpha A_\beta - \partial_\beta
    A_\alpha - i [A_\alpha, A_\beta] \right) \left( \partial_\alpha A_\beta - \partial_\beta
    A_\alpha - i [A_\alpha, A_\beta] \right) + W[A]\,,
\end{equation}
we can rewrite the ERG equation as
\begin{equation}
  \left( \beta (g) \partial_g + \int_p \left( - p \cdot \partial_p - D
      + 1 \right) A_\mu^a (p) \cdot \frac{\delta}{\delta A_\mu^a (p)}
    - \mathcal{D} \right) e^{W_g [A]} = 0
\end{equation}
where
\begin{equation}
  \beta (g) = - \frac{1}{(4 \pi)^2} \frac{4}{3} g^2
\end{equation}
is the 1-loop beta function.

\section{Conclusions}

In this paper we have discussed the multiple products of current
operators using the exact renormalization group (ERG) formalism.  The
multiple products are characterized by two mutually consistent
equations: one is the ERG differential equation and the other is the
Ward-Takahashi (WT) identity.  We have argued that these two equations
suffer changes due to the short-distance singularities of the
products, and the revised equations are given by (\ref{sec5-Wt-ERG})
for ERG and (\ref{sec5-Wt-WT}) for the WT identity.  In Sec.~VI we
have calculated the multiple products explicitly by solving these
equations for the Gaussian fixed-point.  The guiding principle in
these calculations is the locality of the operators.  Since the
momenta below the cutoff have not been integrated, the coefficient
functions for the products of the current are analytic at zero
momenta.

There are some future directions we can consider.  We may consider a
theory such as QCD with fields other than the chiral fermions.  Or we
may consider a more nontrivial fixed-point Wilson action.  We also
think it interesting to study the multiple products of other composite
operators such as the energy-momentum tensor.

\appendix

\section{Invariance of the Wilson action} \label{appendix-invariance-action}

Given a Wilson action $S [\psi, \bar{\psi}]$, its invariance under
global flavor transformations is most straightforwardly given by
\begin{equation}
  \int_p \left[ \bar{\psi} (-p) T^a \Ld{\bar{\psi} (-p)} S_t  - S_t
    \Rd{\psi (p)} T^a \psi (p) \right] = 0\,.\label{appendix-invariance}
\end{equation}
We wish to show that this is equal to (\ref{sec2-symmetry-Wilson}) which is
\begin{equation}
\E^a (0) \equiv  \int_p K(p) \Tr \left[ \Ld{\bar{\psi} (-p)} \left( \bar{\Psi} (-p) T^a
        e^{S_t}\right)
      - \left( e^{S_t} T^a \Psi (p)\right) \Rd{\psi (p)} \right] =
    0\,,\label{appendix-symmetry-Wilson}
\end{equation}
where
\begin{subequations}\label{appendix-Psi-Psibar}
\begin{align}
  \Psi (p) &= \frac{1}{K(p)} \left( \psi (p) + h_F (p) \Ld{\bar{\psi}
             (-p)} S_t \right)\,,\\
  \bar{\Psi} (-p) &= \frac{1}{K(p)} \left( \bar{\psi} (-p) + S_t
                    \Rd{\psi (p)} h_F (p) \right)\,.
\end{align}
\end{subequations}
Substituting (\ref{appendix-Psi-Psibar}) into (\ref{appendix-symmetry-Wilson}), we
obtain
\begin{align}
\E^a (0) &= \int_p \left[ - \left( \bar{\psi} (-p) + S_t \Rd{\psi (p)} h_F (p)
  \right) T^a \Ld{\bar{\psi} (-p)} S_t\right.\nt\\
  &\left.\qquad + S_t \Rd{\psi (p)} T^a \left( \psi (p) + h_F (p)
    \Ld{\bar{\psi} (-p)} S_t \right) \right]\nt\\
& \quad + \int_p  \Tr \left[ \delta (0) T^a - \delta (0) T^a
                                                    \right]\nt\\
  &\quad + \int_p \Tr \left[\Ld{\bar{\psi} (-p)} S_t \Rd{\psi (p)} h_F (p) T^a - T^a
                                                    h_F (p) \Ld{\bar{\psi} (-p)} S_t \Rd{\psi (p)} \right]\nt\\
  &= \int_p \left[ - \bar{\psi} (-p) T^a \Ld{\bar{\psi} (-p)} S_t +
    S_t \Rd{\psi (p)} T^a \psi (p) \right] = 0\,,
\end{align}
which is (\ref{appendix-invariance}).

\section{Universal Cutoff Integrals}\label{appendix-integrals}

We give four integrals involving a cutoff function $K (p)$.  The
values of these integrals are universal in the sense that they do not
depend on the choice of $K(p)$ as long as $K (0) = 1$ and $K (p)$
vanishes asymptotically as $p^2 \to \infty$.  The functions $h$ and
$f$ are defined by
\begin{subequations}
  \begin{align}
    h(p) &\equiv \frac{1 - K(p)}{p^2}\,,\\
    f (p) &\equiv \left( p \cdot \partial_p + 2 \right) h (p) =
            \frac{\Delta (p)}{p^2}\,,\\
    \Delta (p) &\equiv - p \cdot \partial_p K(p)\,.
  \end{align}
\end{subequations}

\subsection{}

For $n = 0, 1, 2, \cdots$, we obtain
\begin{align}
  \int_q f (q) h (q) \left( q^2 h(q) \right)^n
  &= \int_q \left( q \cdot \partial_q + 2 \right) h(q) \cdot
    \frac{1}{q^2} \left( q^2 h(q)\right)^{n+1}\nt\\
  &= \int_q \frac{1}{q^4} q \cdot \partial_q \lb \frac{\left(q^2
    h(q)\right)^{n+2}}{n+2}\rb\nt\\
  &= \frac{2 \pi^2}{(2\pi)^4} \int_0^\infty dq^2 \frac{d}{dq^2} \lb \frac{\left(q^2
    h(q)\right)^{n+2}}{n+2}\rb\nt\\
  &= \frac{1}{(4 \pi)^2} \frac{2}{n+2}\,.
\end{align}

\subsection{}

For $n = 0, 1, 2, \cdots$, we obtain
\begin{align}
  \int_q K(q) h(q) \left( q^2 h(q)\right)^n \left( h(q) + q^2 h' (q)
  \right)
  &= \int_q \left(1 - q^2 h(q)\right) h(q) \left(q^2 h(q)\right)^n
    \frac{d}{dq^2} \left( q^2 h(q)\right)\nt\\
  &= \int_q \frac{1}{q^2} \left( 1 - q^2 h(q)\right) \left(q^2
    h(q)\right)^{n+1} \frac{d}{dq^2} \left( q^2 h(q)\right)\nt\\
  &= \frac{1}{(4 \pi)^2} \int_0^\infty dq^2 \frac{d}{dq^2} \left(
    \frac{\left(q^2 h(q)\right)^{n+2}}{n+2} - \frac{\left(q^2
    h(q)\right)^{n+3}}{n+3} \right)\nt\\
  &= \frac{1}{(4 \pi)^2} \frac{1}{(n+2)(n+3)}\,.
\end{align}

\subsection{}

\begin{align}
  \int_q f(q) \left( q^2 h' (q) + \frac{1}{3} q^4 h'' (q) \right)
  &= \frac{1}{(4 \pi)^2} \int_0^\infty x dx\, \underbrace{f(x)}_{= 2
    \left( x \frac{d}{dx} + 1 \right) h(x)} \left( x
    \frac{d}{dx} + \frac{1}{3} x^2 \frac{d^2}{dx^2} \right) h(x)\nt\\
  &= \frac{2}{(4 \pi)^2} \int_0^\infty dx\, x \left(1 + x
    \frac{d}{dx}\right) h(x) \cdot x \left( \frac{d}{dx} + \frac{1}{3}
    x \frac{d^2}{dx^2} \right) h (x)\nt\\
  &= \frac{2}{(4 \pi)^2} \int_0^\infty dx\, \frac{d}{dx} \left(
    \frac{1}{3} x^3 h (x) h' (x) + \frac{1}{6} x^4 h' (x)^2 \right)\nt\\
  &=  \frac{2}{(4 \pi)^2} \left( - \frac{1}{3} + \frac{1}{6} \right)
    = - \frac{1}{(4 \pi)^2} \frac{1}{3}\,.
\end{align}

\subsection{}

\begin{align}
&  \int_q K(q) \left( h' (q) + q^2 h'' (q) + \frac{1}{6} q^4 h''' (q)
  \right)\,\nt\\
  &\quad= \frac{1}{(4 \pi)^2} \int_0^\infty dq^2 \, q^2 K(q) \left( h' (q)
    + q^2 h'' (q) + \frac{1}{6} h'' (q) \right)\,\nt\\
  &\quad= \frac{1}{(4 \pi)^2} \int_0^\infty dq^2 \, \frac{d}{dq^2} \left[ -
    \frac{1}{6} q^4 K(q) K'' (q) + \frac{1}{6} \left( q^4 \frac{1}{2}
    K' (q)^2 - q^2 K(q) K' (q) \right) + \frac{1}{12} K(q)^2 \right]\,\nt\\
  &\quad= - \frac{1}{(4 \pi)^2} \frac{1}{12}\,.
\end{align}

\subsection{Check of (\ref{sec6-integral})}

We wish to check (\ref{sec6-integral}) in sec.~6, which can be written
as
\begin{equation}
  \int_q \left( f(q) h (q) q^2 - K(q) (2 h (q) + f (q)) \right) = 0\,.
\end{equation}
The integrand is a total derivative:
\begin{align}
  \left( q \cdot \partial_q + 4 \right) \left( K(q) h (q) \right)
  &= - \Delta (q) h (q) + K (q) f (q) + 2 K(q) h(q)\nt\\
  &= - q^2 f(q) h(q) + K(q) \left(2 h (q) + f (q)\right)\,.
\end{align}
Since $K(q) h(q)$ vanishes at $q^2=0\,, \infty$, the integral vanishes.

\section{Corrections to the ERG equation and the WT identity}\label{appendix-equations-for-J}

Differentiating (\ref{sec5-Wt-ERG}) and (\ref{sec5-Wt-WT}) with respect
to the source $A$, we obtain the ERG equation and the WT identity
for the products of current operators.

\subsection{Product of Two}

The ERG differential equation is
\begin{equation}
  \left(\partial_t + p_1 \cdot \partial_{p_1} + p_2 \cdot
  \partial_{p_2} + 2 - \mathcal{D}_t \right) \op{J_\alpha^a (p_1) J_\beta^b (p_2)}
= b(t) \delta (p_1+p_2) \,\delta^{ab} \left(
    p_{1\alpha} p_{1\beta} - p_1^2 \delta_{\alpha\beta}\right)\,.
\end{equation}

The WT identity has no anomaly.
\begin{equation}
  p_\alpha \op{J_\alpha^a (p) J_\beta^b (p_1)}
  = i f^{abc} J_\beta^c (p+p_1) + \E^a (p) \star J_\beta^b (p_1)\,.
\end{equation}

\subsection{Product of Three}

ERG differential equation
\begin{align}
&  \left( \sum_{i=1}^3 p_i \cdot \partial_{p_i} + 3 - \mathcal{D}_t \right)
   \op{J_\alpha^a (p_1) J_\beta^b (p_2) J_\gamma^c
    (p_3)}\nt\\
  &= b(t) \Bigg[ \delta \left( \sum_{i=1}^3
    p_i \right) \Tr T^a \left[ T^b , T^c\right] \, \lb \delta_{\alpha\beta} (p_1-p_2)_\gamma +
    \delta_{\beta\gamma} (p_2-p_3)_\alpha + \delta_{\gamma\alpha}
    (p_3-p_1)_\beta \rb\nt\\
  &\qquad\qquad - \lb  \delta (p_1+p_2)
    \left( p_{1 \alpha} p_{1 \beta} - p_1^2 \delta_{\alpha\beta}
    \right) \delta^{ab} J_\gamma^c (p_3) + \delta (p_2 + p_3) \left(
    p_{2 \beta} p_{2\gamma} - p_2^2 
    \delta_{\beta\gamma} \right) \delta^{bc} J_\alpha^a (p_1) \right.\nt\\
  &\left.\qquad\qquad\quad  + \delta (p_3+p_1)
    \left( p_{3 \gamma} p_{3 \alpha} - p_3^2 \delta_{\gamma\alpha}
    \right) \delta^{ca} J_\beta^b (p_2) \rb \Bigg]\,.
\end{align}

The WT identity can be anomalous:
\begin{align}
  p_\alpha \op{J_\alpha^a (p) J_{\beta}^b (q) J_\gamma^c (r)}
  &= i f^{abd} \op{J_\beta^d (p+q) J_\gamma^c (r)} + i f^{acd}
    \op{J_\beta^b (q) J_\gamma^d (p+r)}\nt\\
  &\quad + \E^a (p) \star \op{J_\beta^b (q) J_\gamma^c (r)}\nt\\
  &\quad - \frac{\mathcal{A}}{2} \delta \left(p+q+r\right)
    \Tr T^a \lb T^b, T^c\rb\, \ep_{\alpha\beta\gamma\delta} p_\alpha (q-r)_\delta\,.
\end{align}

\subsection{Product of Four}

ERG differential equation
\begin{align}
&  \left( \sum_{i=1}^4 p_i \cdot \partial_{p_i} + 4 -
  \mathcal{D}_t\right) \op{J_\alpha^a (p_1) J_\beta^b (p_2) J_\gamma^c
  (p_3) J_\delta^d (p_4)}\nt\\
  &= b(t) \Bigg[\, \delta (p_1+p_2+p_3+p_4)
    \left[ \Tr T^a T^b T^c T^d \left(
  \delta_{\alpha\beta} \delta_{\gamma\delta} + \delta_{\beta\gamma}
    \delta_{\delta\alpha} - 2 \delta_{\alpha\gamma}
    \delta_{\beta\delta} \right)\right.\nt\\
  &\qquad\quad+ \Tr T^a T^b T^d T^c \left( \delta_{\alpha\beta}
    \delta_{\delta\gamma} + \delta_{\beta\delta} \delta_{\gamma\alpha}
    - 2 \delta_{\alpha\delta} \delta_{\beta \gamma} \right)
    + \Tr T^a T^c T^b T^d \left( \delta_{\alpha\gamma}
    \delta_{\beta\delta} + \delta_{\gamma\beta} \delta_{\delta\alpha}
    - 2 \delta_{\alpha\beta} \delta_{\gamma\delta} \right)\nt\\
  &\qquad\quad + \Tr T^a T^c T^d T^b \left( \delta_{\alpha\gamma}
    \delta_{\delta\beta} + \delta_{\gamma\delta} \delta_{\beta\alpha}
    - 2 \delta_{\alpha\delta} \delta_{\gamma\beta} \right)
    + \Tr T^a T^d T^b T^c \left( \delta_{\alpha\delta}
    \delta_{\beta\gamma} + \delta_{\delta\beta} \delta_{\gamma\alpha}
    - 2 \delta_{\alpha\beta} \delta_{\delta\gamma} \right)\nt\\
  &\qquad\quad \left. + \Tr T^a T^d T^c T^b \left( \delta_{\alpha\delta}
    \delta_{\gamma\beta} + \delta_{\delta\gamma} \delta_{\beta\alpha}
    - 2 \delta_{\alpha\gamma} \delta_{\delta\beta} \right) \right]\nt\\
  &\qquad+  \delta (p_1+p_2+p_3) \Tr T^a \left[ T^b, T^c\right] \lb
    \delta_{\alpha\beta} (p_1-p_2)_\gamma + 
    \delta_{\beta\gamma} (p_2-p_3)_\alpha + \delta_{\gamma\alpha}
    (p_3-p_1)_\beta \rb J_\delta^d (p_4)\nt\\
  &\qquad + \delta (p_1+p_2+p_4) \Tr T^a \left[ T^b, T^d\right] \lb \delta_{\alpha\beta}
    (p_1-p_2)_\delta + \delta_{\beta\delta} (p_2-p_4)_\alpha +
    \delta_{\delta\alpha} (p_4-p_1)_\beta \rb J_\gamma^c (p_3)\nt\\
  &\qquad + \delta (p_1+p_3+p_4)  \Tr T^a \left[ T^c, T^d\right] \lb
    \delta_{\alpha\gamma} (p_1-p_3)_\delta + \delta_{\gamma\delta}
    (p_3-p_4)_\alpha + \delta_{\delta\alpha} (p_4-p_1)_\gamma \rb
    J_\beta^b (p_2)\nt\\
  &\qquad + \delta (p_2+p_3+p_4) \Tr T^b \left[ T^c, T^d\right]
    \lb \delta_{\beta\gamma} (p_2-p_3)_\delta + \delta_{\gamma\delta}
    (p_4-p_4)_\beta + \delta_{\delta\beta} (p_4-p_2)_\gamma \rb
    J_\alpha^a (p_1) \nt\\
  &\qquad + \delta (p_1+p_2) \left( p_1^2 \delta_{\alpha\beta} -
    p_{1\alpha} p_{1\beta} \right) \delta^{ab} \op{J_\gamma^c (p_3) J_\delta^d
    (p_4)} + \delta (p_1+p_3) \left( p_1^2
    \delta_{\alpha\gamma} - p_{1\alpha} p_{1\gamma} \right)
    \delta^{ac} \op{J_\beta^b (p_2) J_\gamma^d (p_4)} \nt\\
  &\qquad+ \delta (p_1+p_4) \left( p_1^2 \delta_{\alpha\delta} - p_{1\alpha} p_{1\delta} \right)
  \delta^{ad} \op{J_\beta^b (p_2) J_\gamma^c  (p_3)}
 + \delta (p_2+p_3) \left( p_2^2 \delta_{\beta\delta} -
    p_{2\beta} p_{2\delta} \right) \delta^{bc} \op{J_\alpha^a (p_1) J_\delta^d   (p_4)}\nt\\
&\qquad  + \delta (p_2 + p_4) \left( p_2^2 \delta_{\beta\delta} -
    p_{2\beta} p_{2\delta} \right) \delta^{bd} \op{J_\alpha^a (p_1) J_\gamma^c (p_4)}
 +  \delta (p_3+p_4) \left(  p_3^2 \delta_{\gamma\delta}
       -  p_{3\gamma} p_{3\delta} \right)  \delta^{cd} \op{J_\alpha^a (p_1)
 J_\beta^b (p_2)}  \Bigg]\,.
\end{align}
The WT identity can be anomalous:
\begin{align}
&  p_\alpha \op{J_\alpha^a (p) J_\beta^b (q) J_\gamma^c (r) J_\delta^d
  (s)}\nt\\
  &= i f^{abe} \op{J_\beta^e (q+p) J_\gamma^c (r) J_\delta^d (s)}
    + i f^{ace} \op{J_\beta^b (q) J_\gamma^e (r+p) J_\delta^d (s)}
 + i f^{ade} \op{J_\beta^b (q) J_\gamma^c (r) J_\delta^e (s+p)}\nt\\
&\quad + \E^a (p) \star \op{ J_\beta^b (q) J_\gamma^c (r) J_\delta^d
    (s)}\nt\\
  & \quad- \frac{\mathcal{A}}{2} \Big[ \delta (p+q+r+s)\, p_{1\alpha}
    \ep_{\alpha\beta\gamma\delta} \Tr T^a \left( T^b \op{T^c,
    T^d}+T^c\op{T^d, T^b} + T^d \op{T^b, T^c} \right)\nt\\
  &\qquad\quad +     \delta (p+q+r) \Tr T^a \lb T^b, T^c \rb \ep_{\alpha\beta\gamma\ep} p_\alpha
    (q-r)_\ep J_\delta^d (s)\nt\\
   &\qquad\quad + \delta (p+q+s) \Tr T^a \lb T^b, T^d \rb \ep_{\alpha\beta\delta\ep} p_\alpha
    (q-s)_\ep J_\gamma^c (r)\nt\\
  &\qquad\quad + \delta (p+r+s) \Tr T^a \lb T^c, T^d \rb
    \ep_{\alpha\gamma\delta\ep} p_\alpha  (r-s)_\ep J_\beta^b (q) \Big]\,.
\end{align}

\begin{acknowledgments}
  I would like to thank Prof. P.~D.~Prester of the University of
  Rijeka, Croatia for raising a question that gave me a motivation for this work.
\end{acknowledgments}

\bibliography{paper}

\end{document}